\documentclass{ws-ijmpa}
\usepackage[super,compress]{cite}
\usepackage{physics}
\usepackage{color}
\usepackage{here}
\usepackage{siunitx}
\usepackage{todonotes}
\usepackage{algorithm,algorithmic}
\usepackage{graphicx}
\usepackage{hyperref}
\usepackage{multirow}
\usepackage{booktabs}
\newcommand{\NC}{N}

\def\draftnote{\today\quad\currenttime\quad WSPC/INSTRUCTION FILE\qquad\jobname\\\hspace*{14.0cm}\raisebox{1em}[0em][0em]{HUPD-2208}\hspace*{-3.51cm}\raisebox{-0em}[0em][0em]{IFT-UAM/CSIC-22-78}}

\begin{document}
\markboth{Antonio~Gonz\'{a}lez-Arroyo, Ken-Ichi~Ishikawa, Yingbo~Ji, and Masanori~Okawa}%
{Perturbative study of large $\NC$ principal chiral model with twisted reduction}

%%%%%%%%%%%%%%%%%%%%% Publisher's Area please ignore %%%%%%%%%%%%%%%
%
\catchline{}{}{}{}{}
%
%%%%%%%%%%%%%%%%%%%%%%%%%%%%%%%%%%%%%%%%%%%%%%%%%%%%%%%%%%%%%%%%%%%%

\title{Perturbative study of large $\NC$ principal chiral model with twisted reduction}

\author{Antonio~Gonz\'{a}lez-Arroyo}
\address{Instituto de F\'{i}sica Te\'{o}rica UAM-CSIC, Nicol\'{a}s Cabrera 13-15, Universidad Aut\'{o}noma de Madrid, Cantoblanco, E-28049, Madrid, Spain;\\
         Departamento de F\'{i}sica Te\'{o}rica, M\'{o}dulo 15, Universidad Aut\'{o}noma de Madrid, Cantoblanco, E-28049, Madrid, Spain}

\author{Ken-Ichi~Ishikawa$^{\dag\ddag}$\footnote{Email \textit{ishikawa@theo.phys.sci.hiroshima-u.ac.jp}}, Yingbo~Ji$^{\ddag}$\footnote{Corresponding author,  email \textit{d202186@hiroshima-u.ac.jp}}, and Masanori~Okawa$^{\ddag}$}
\address{$^{\dag}$Core of Research for the Energetic Universe, Graduate School of Advanced Science and Engineering, Hiroshima University, Higashi-Hiroshima, Hiroshima 739-8526, Japan\\
         $^{\ddag}$Graduate School of Advanced Science and Engineering, Hiroshima University, Higashi-Hiroshima, Hiroshima 739-8526, Japan}

\maketitle

\begin{history}
\received{Day Month Year}
\revised{Day Month Year}
\end{history}

\begin{abstract}

We compute the first four perturbative coefficients of the internal energy for the
twisted reduced principal chiral model (TRPCM) using numerical
stochastic perturbation theory (NSPT). This matrix model has 
the same large $\NC$ limit as the ordinary principal chiral model
(PCM) at infinite volume. Indeed, we verify that the first three
coefficients match the analytic result for the PCM coefficients at
large $\NC$ with a precision of three to four significant digits. The fourth
coefficient also matches our own NSPT calculation of the corresponding
PCM coefficient at large $\NC$. The finite-$\NC$
corrections to all coefficients beyond the leading order are smaller
for TRPCM than for PCM. We analyze the variance to determine the feasibility of extending the calculations to higher orders.
\keywords{Lattice Field Theory; Large $\NC$; Perturbation Theory.}
\end{abstract}

\ccode{PACS numbers:11.10.-z,12.39.Fe,11.10.Ef,11.25.Db}

%\tableofcontents

\newcommand{\convolution}{\circledast}

\section{Introduction}

The two dimensional $SU(\NC)\times SU(\NC)$ principal chiral model (PCM) is often considered as a toy model for four-dimensional pure gauge theory, because they share many interesting properties~\cite{PCMandGT2,PCMandGT,Shigemitsu:1980tx,Rossi:1996hs,Green:1980bg,Abdalla:1984iq,Wiegmann:1984ec,Green:1981tr,Campostrini:1994sh,Campostrini:1994ih}, such as generating a mass gap~\cite{Balog:1992cm} and being asymptotically free. To study these properties, various perturbative and non-perturbative methods have been applied. The large $\NC$ expansion method can be applied to the PCM and gauge theories, where the standard weak-coupling expansion is reorganized to obtain better behavior and to simplify the Feynman diagrams to explain the perturbative properties of the theories. The lattice method has been applied to these models to study non-perturbative properties, and combining the lattice method with large $\NC$ expansion is a natural way to understand their origin. Exploring the lattice PCM at large $\NC$ is an important step toward the understanding  of four-dimensional lattice gauge theories.

Our interest is in studying the high-order perturbative behavior of the lattice PCM at the large $\NC$ limit. Motivated by the recently developed resurgence theory~\cite{Dunne:2014bca,Dorigoni:2014hea}, 
one can deduce the non-perturbative effects, such as renormalon~\cite{Pazarbasi:2019web,Beneke:1998ui} and the complex saddle point of the action, from the perturbation series in quantum field theory and quantum mechanics~\cite{Gozzi:2020wef}.
To extract the non-perturbative effects from the perturbation series, we need to investigate the
behavior of the coefficients in terms of the coupling $g$ at rather high orders, such as $\order{g^{40}}-\order{g^{100}}$. For models that are not analytically solvable, this is a difficult task without the help of numerical methods.

Numerical stochastic perturbation theory (NSPT) is a numerical method that can extract the perturbative coefficients at high orders semi-automatically~\cite{DiRenzo:1994av}. Recently, Bruckmann and Puhr have successfully extracted the renormalon behavior of the PCM from the high-order behavior of
the perturbation coefficients of the internal energy~\cite{Bruckmann:2019mky,Puhr:2018zmt}.
They computed the coefficients up to $\order{\lambda^{20}}$, where $\lambda$ is the 't Hooft coupling,
for $N=3$--$12$ on several lattice sizes up to $48^2$ to extract the infinite-volume limits.
The success of the high-order NSPT on the two-dimensional PCM in the large $N$ limit
might lead to a similar one for the four-dimensional large $N$ gauge theories. However,
applying the strategy of taking both the large $N$ and infinite volume limits to the four-dimensional gauge theories
could be computationally quite expensive.
One approach to tackle the large $N$ limit without taking the infinite volume limit
is using the volume reduction method which has been first applied to the pure lattice gauge theory by Eguchi and Kawai~\cite{Eguchi:1982nm}.

Eguchi and Kawai showed
that as $N\rightarrow \infty$ for the pure lattice gauge theory, the structure of the lattice could collapse into a single point
provided that the $Z_N$ symmetry is maintained, which led the Eguchi-Kawai-volume-reduced (EK) model.
However, it has been realized that the original EK model does not reproduce the proper large $N$ limit
because of the spontaneous breaking of the $Z_N$ symmetry in the continuum limit~\cite{Bhanot:1982sh}.
Two of the present authors extended it to the twisted Eguchi-Kawai (TEK) model by introducing twisted boundary conditions,
by which the internal degree of colour index is explicitly related to the degree of space-time lattice sites, and
have shown that it realizes the correct continuum limit non-perturbatively~\cite{Gonzalez-Arroyo:1982hwr,Gonzalez-Arroyo:1982hyq,GonzlezArroyo2010}.
Twisted boundary conditions can be introduced for the PCM, and the  corresponding twisted reduced principal chiral model (TRPCM) has been first investigated in  Ref.~\citen{Gonzalez-Arroyo:1984agg}. 
The large $\NC$ limit of the TRPCM was later investigated in Ref.~\citen{GonzlezArroyo2018} non-perturbatively,
where the choice of the flux value of the twisted boundary condition is extensively studied.
By combining the TRPCM and NSPT, we can study the high-order perturbation coefficients in the large $\NC$ limit 
without taking explicitly infinite volume limit, as the large $\NC$ limit involves implicitly taking the infinite volume limit.
Another important property of the large $\NC$ limit of the twisted reduced models is the realization of the master 
field property, a short simulation is enough to obtain the expectation values of observables, with the reduction of the variance through the factorization property in the large $\NC$ limit.
A study in this direction for the pure gauge theory has started in Ref.~\citen{GonzlezArroyo2019}.

In this paper, we study the large $\NC$ behavior of the perturbative coefficients of the internal energy for the TRPCM using NSPT. 
Before studying much higher orders in the large $\NC$ limit, we first study the first fourth coefficients using NSPT. We examine the feasibility of a short NSPT simulation combined with the large $\NC$ factorization and the master field property at a fixed but sufficiently large $\NC$.
NSPT is a numerical application of the stochastic quantization method~\cite{Parisi:1980ys,Damgaard:1987rr} to the lattice models.
Recently, it has a lot of achievements in many fields, including perturbative calculation in full and 
quenched QCD~\cite{DiRenzo:1994sy,Brambilla:2013sba,DallaBrida:2013thf}, 
high-order perturbative behavior~\cite{Bali:2013pla,DelDebbio:2018ftu,Bali:2014fea,Puhr:2018zmt,Bruckmann:2019mky}
always recognized as renormalon and even $g-2$ in QED~\cite{Kitano:2021ecc}. 
The original NSPT~\cite{DiRenzo:1994av} was based on the Langevin equation, and it has been extended 
to the hybrid molecular dynamics (HMD) based NSPT~\cite{DallaBrida:2013thf,DallaBrida:2017pex}.
We use the HMD-based NSPT for the TRPCM in this paper.

The present article is organized as follows. 
In Sec.~\ref{sec:ModelAndNSPT} we will briefly review the cornerstone of our study, including reviews on the two dimensional $\text{SU}(\NC)$ TRPCM and the HMD-based NSPT algorithm.
In Sec.~\ref{sec:NSPTresults} we present our numerical results obtained by applying NSPT to our model.
We explore the $\NC$ dependence of the first fourth-order coefficients of the internal energy.
We also conduct a qualitative survey 
on the relation between $\NC$ dependence of the variance and the statistical error through the factorization property, from which the number of independent samples required for a fixed statistical error is evaluated.
In Sec.~\ref{sec:outlook} we combine the results obtained in Sec.~\ref{sec:NSPTresults}, the finite $\NC$ corrections and the $\NC$ dependence of the statistical error,
to estimate the number of independent samples we need for performing high-order simulations at a single but sufficiently large $\NC$. 
Finally, a short summary is included in the last section.

\section{Twisted reduced principal chiral model and numerical stochastic perturbation theory}
\label{sec:ModelAndNSPT}

In this section, we briefly describe the twisted reduced principal chiral model (TRPCM),
and the numerical stochastic perturbation theory (NSPT) and its application to the TRPCM.
We employ the hybrid molecular dynamics type NSPT algorithm including a randomized trajectory length scheme 
to circumvent the ergodicity problem in perturbation theory.

\subsection{Review of TRPCM}
The two dimensional $\mathrm{SU}(\NC)$
twisted reduced principal chiral model (TRPCM)~\cite{Das:1983jv,Gonzalez-Arroyo:1984agg} 
is specified by the following partition function:
\begin{align}
       Z&= \int \dd{U} \exp\qty{-S[U]}, \notag\\
   S[U] &= b \NC \sum_{\mu=1,2} \Tr\qty(V_\mu  V_\mu^{\dag}),
\label{eq:PartitionFunction}\\
  V_\mu &\equiv \Gamma_{\mu} U \Gamma_{\mu}^{\dag}-U,\notag
\end{align}
where $U$ is a $\mathrm{SU}(\NC)$ matrix, $b$ is the inverse 't~Hooft coupling defined as $b = 1/(\NC g^2)$, and 
$\Gamma_\mu$'s are the twist matrices.
This model is deduced by the twisted volume reduction method
from the lattice 
$\mathrm{SU}(\NC)\times \mathrm{SU}(\NC)/Z_{\NC}$ principal chiral model (PCM) with twisted boundary conditions~\cite{Das:1983jv,Gonzalez-Arroyo:1984agg}.
In the two dimensional case, the $\Gamma_\mu$ matrices satisfy
't~Hooft algebra\cite{GonzlezArroyo2018}:
\begin{align}
  \Gamma_{1} \Gamma_{2}=\exp \qty{i\frac{2 \pi K}{\NC}} \Gamma_{2} \Gamma_{1} \quad \Gamma_\mu \in \mathrm{SU}(\NC).   
\label{eq:TWISTMAT}
\end{align}
For a given $\NC$ and flux $K$, specific  examples of two matrices, $\Gamma_\mu$ are provided by choosing a permutation matrix and a clock matrix according to the following equation respectively.
\begin{align}
 (\Gamma_1)_{i,j} & = \delta_{\mathrm{mod}}(i,\NC)+1,j,\\
 (\Gamma_2)_{i,j} & = \delta_{i,j}\exp\qty{i \frac{(2j-1-\NC)\pi K}{\NC}}.
\end{align}
The TRPCM,  Eq.~\eqref{eq:PartitionFunction}, has the following global-symmetry:
\begin{alignat}{3}
  U &\to \Gamma(n) U \Gamma(n)^{\dag}, &\quad& \Gamma(n) \equiv \qty(\Gamma_1)^{n_1} \qty(\Gamma_{2})^{n_2}, \quad (n_1,n_2 = 0,\cdots,\NC-1),\\
  U &\to U Z = Z U, &\quad& Z \in Z_{\NC}.
\label{eq:CENTERZN}
\end{alignat}

For large $\NC$, the planar sector  of the TRPCM effectively corresponds to the lattice PCM defined on a square
lattice with the size $L = \NC$.
Therefore, as $\NC \rightarrow \infty$, the TRPCM meets both large $\NC$ limit and infinite volume limit simultaneously, and 
eventually coincides with the PCM in both the large $\NC$ and infinite volume limits.

The choice of $K$, the flux parameter, is important to realize the coincidence non-perturbatively 
in the large $\NC$ and the continuum limit, and has been investigated in Refs.~\citen{GonzlezArroyo2010,GonzlezArroyo2018,GonzlezArroyo2019}. 
For the equivalence in the large $\NC$ limit, the center-symmetry \eqref{eq:CENTERZN} has to be maintained.
Taking the large $\NC$ limit with a constant $K$, 
the twist phase $\exp \qty{\frac{2 \pi i K}{\NC}}$ in Eq.~\eqref{eq:TWISTMAT}  might become ineffective to prevent the breaking of the center-symmetry\cite{Profumo2002,GonzlezArroyo2018,Chamizo:2016msz}.
To avoid that, a feasible proposal of choosing the relation between $K$ and $\NC$ is given in Refs.~\citen{GonzlezArroyo2018,Chamizo:2016msz}. 
 For fixed $N$, $K$ is chosen such that
\begin{align}
   \min_{e} e\norm{\dfrac{Ke}{\NC}} > \Lambda,
\label{eq:ratio}
\end{align}
where $e$  runs over all integers co-prime with $\NC$, $\norm{\cdot}$ is the distance to the nearest integer, 
and $\Lambda$ is a threshold of $\Lambda \simeq 0.15$. 
We follow this choice in NSPT.
It is also known that the twist phase is important for eliminating non-planar diagrams in the large
$\NC$ limit in perturbation theory, leading to smaller finite $\NC$ corrections.

We employ the internal energy density operator $E$ defined by
\begin{equation}
    E = \frac{1}{2\NC} \sum_\mu \Re\Tr\qty[U\Gamma_\mu U^\dag \Gamma_\mu^\dag],
\label{eq:defintE}
\end{equation}
as the main observable in this paper.

\subsection{A brief introduction to NSPT for TRPCM}

NSPT is a powerful tool for estimating high-order perturbative coefficients in quantum mechanics and lattice field theory. 
Since it builds on top of the stochastic quantization, the original NSPT is based on the Langevin equation~\cite{DiRenzo:1994av}. 
However, some limitations, including the lack of a high-order integrator to make the systematic 
error under control and the critical slow down in a critical situation, occur in the Langevin-based NSPT.

Because of the disadvantages of the Langevin-based NSPT, in this paper, we employ a hybrid molecular dynamics (HMD) based NSPT~\cite{4th,4th2}. The HMD method has been utilized to study a variety of complicated and non-local systems non-perturbatively and,
in particular, its derivative hybrid Monte Carlo (HMC) has gained an important place in numerical computation and statistics.
Applying the HMD to NSPT is a natural way to improve the efficiency of NSPT~\cite{DiRenzo:1994av}. 
A further improvement of NSPT beyond the HMD has been studied in Ref.~\citen{4th}.

To derive NSPT for the TRPCM more naturally, we first introduce the non-perturbative
HMD simulation for the TRPCM as a starting point. 
The partition function for the HMD based Monte Carlo algorithm is translated from \eqref{eq:PartitionFunction} as
\begin{align}
Z&=\int \dd{U} \dd{P} e^{-H[P,U]}, \\
H[P,U] &= \frac{1}{2} \mathrm{Tr}\left[P^2\right] +S[U].
\end{align}
where $P$ is a $\NC\times \NC$ traceless Hermitian matrix. 
The variable $\{P,U\}$ is stochastically generated to satisfy the probability density $\dd{U} \dd{P} e^{-H[P,U]}/Z$ 
in the HMD algorithm. 
The Markov chain for the density is constructed by regarding $H[P,U]$ as a Hamiltonian and introducing 
a fictitious time $t$ for the dynamical variables $\{P,U\}$.
The equation of motion for $\{P,U\}$ is 
\begin{align}
\dv{U}{t} &= i P U, \quad 
\dv{P}{t}  =  F,
\label{eq:HMDEOM}
\\
F &= i b \NC \qty(V - \frac{1}{\NC}\Tr\qty[V]), \quad 
V  = UX-(UX)^\dag,\\
X & = \sum_\mu\qty[\Gamma_\mu^\dag U^\dag \Gamma_\mu + \Gamma_\mu U^\dag \Gamma_\mu^\dag ].
\end{align}

The HMD algorithm uses a symplectic integration scheme to approximate the time
evolution of the equation of motion, and the variable $P$ is periodically refreshed 
as a stochastic variable from the Gaussian distribution $\exp[-\Tr\qty[P^2]/2]$.  The Markov Chain Monte Carlo sampling of the HMD algorithm is used for the periodic
sampling from the trajectory of $\{P,U\}$ as a function of time $t$.

NSPT formulation is obtained by replacing the variables $\{P,U\}$ in the EoM of the non-perturbative HMD \eqref{eq:HMDEOM}
with their perturbative expansion in terms of the coupling constant. 
To take the large $\NC$ limit of the TRPCM, the 't~Hooft coupling $\lambda \equiv b^{-1}$ is a natural expansion parameter.
The perturbative expansion for $U$ and $P$ is defined by
\begin{align}
  U & = \sum_{k=0}^{\infty} \lambda^{k/2} U^{(k)},
\quad
  P   = \lambda^{-1/2} \sum_{k=1}^{\infty} \lambda^{k/2} P^{(k)},
\end{align}
where $U^{(0)}=I$ and $P^{(0)}=0$ are imposed as the perturbative vacuum and non-dynamical variable.
The first coefficient, $P^{(1)}$, is treated as the source of the randomness and generated from 
the Gaussian distribution $\exp[-\Tr\qty[\qty(P^{(1)})^2]/2]$,
 while the higher order coefficients, $P^{(k)}  (k > 1)$, are reset to zero at the beginning of each trajectory.
As the force $F$ starts at $\order{\lambda^{-1}}$, we re-scale the expansion of $P$ by $\lambda^{-1/2}$ and 
the unit of the fictitious time by
\begin{align}
    t = \lambda^{1/2} t^{\prime},
\end{align}
to expand the EoM. 
We will omit $\prime$ for the re-scaled time unit hereafter.

The building blocks for the symplectic time integration scheme are
\begin{align}
 U^{(k)}(t+\delta t) &= \left[ e^{i P \delta t} \convolution  U(t)\right]^{(k)},
\label{eq:NSPTEoMU}
 \\
 P^{(k)}(t+\delta t) & = P^{(k)}(t) + F^{(k)} \delta t,
\label{eq:NSPTEoMP}
\end{align}
where $\delta t$ is the discretized time step. 
The time of $P$ in \eqref{eq:NSPTEoMU}
  and $F^{(k)}$ in \eqref{eq:NSPTEoMP} can be an arbitrarily time near $t$ depending on the integration scheme in which they are involved.
The symbol $\convolution$ denotes the polynomial convolution product for two matrix polynomials.
NSPT expansion for the matrix exponential $e^{i P \delta t}$ is given in Ref.~\citen{GonzlezArroyo2019}.
The perturbative expansion of the force $F$ is derived as
\begin{align}
F^{(k)} &= i\NC\left(V^{(k)} - \frac{1}{\NC}\mathrm{Tr}\left[V^{(k)}\right]\right),\quad
V^{(k)}  = S^{(k)}-S^{(k)^\dagger}, \\
S^{(k)} &= \qty(U \convolution X)^{(k)},\quad
X^{(k)}  = \sum_\mu\qty[\Gamma_\mu^\dag U^{(k)\dag} \Gamma_\mu + \Gamma_\mu U^{(k)\dag} \Gamma_\mu^\dag]. 
\end{align}    

We also note that the HMD algorithm for NSPT possesses problems with non-ergodicity 
as discussed in Refs.~\citen{Mackenze1989,2001,4th,4th2}.
In order to keep it ergodic, two remedies have been established.
One method is to systematically sample all Fourier
mode of field variables by randomly varying the trajectory length $t$ between samples. The alternative is to adjust the trajectory length to be shorter than a length below which the HMD algorithm begins to resemble the original Langevin algorithm. In this study, we employ a method that combine these two approaches.

For the symplectic integrator for \eqref{eq:NSPTEoMU} and \eqref{eq:NSPTEoMP}, 
we employ the 4th-order Omelyan--Mryglod--Folk (OMF) integrator\cite{4th,4th2}, 
the finite integration error was expected to be proportional to $\delta t^4$.
The randomized trajectory length $t_r$ is determined by
\begin{align}
  t_r =  n_r \delta t,  \quad  \delta t = t / N_{\mathit{md}},
\end{align}
where $t$ is a fixed length, $N_{\mathit{md}}$ is an integer for the number of time steps, 
and $n_r$ is a random number generated as $n_r \leftarrow B(1/2,2(N_{\mathit{md}}-1))+1$
with the binomial distribution $B(p,n)$. 
Thus the averaged trajectory length is $\expval{t_r}=t$. 
All numerical results are computed at several $\delta t$ values and extrapolated to 
$\delta t \to 0$ according to the $\delta t^4$ scaling before analyzing the $\NC$ dependence.

The EoM is evolved for the randomized trajectory length and the perturbative coefficients 
$\{U^{(k)}, k=1,\dots\}$, which are stochastic variables, are sampled as the Monte Carlo ensemble. 
The observables are also expanded perturbatively in $\lambda$, where the perturbative coefficients are 
the function of $\{ U^{(k)}, k=1,\dots\}$.
The internal energy density operator $E$ in Eq.~\eqref{eq:defintE} is expanded as
\begin{align}
      E &= \sum_{k=0}^{\infty} \lambda^{k/2} E^{(k)},\\
E^{(k)} &= \frac{1}{2\NC} \sum_\mu \Re\Tr\qty[\qty(U\Gamma_\mu \convolution U^\dag \Gamma_\mu^\dag)^{(k)}].
\end{align}

The expectation value of the coefficient, $\expval{E^{(k)}}$, is evaluated as the statistical
average on the ensemble for $\{U^{(k)}, k=1,\dots\}$. The perturbation series is truncated in fixed
order in the actual numerical simulation. In this paper, we keep the perturbation series up to $\lambda^{4}$.

\section{NSPT results}
\label{sec:NSPTresults}

The numerical results of the TRPCM's perturbative coefficients which were described in the preceding section, are presented in this section. We determine that the PCM and the TRPCM have
equivalent perturbative coefficients in the large $\NC$ limit. We compare the first four perturbative coefficients of the internal energy of the TRPCM with NSPT to that of the PCM in the large $\NC$ limit.
Once these are shown, we confirm the reliability of our result using one of the important properties of 
large $\NC$ limit - factorization, where we examine the variance of the coefficients converging to zero in the limit.
Additionally, we estimate the number of independent samples at a fixed statistical error because the variance and the relative statistical error have a direct
relation. In order to demonstrate the viability of the high-order computation with a single
brief simulation at a sufficiently large but finite $\NC$, the finite $\NC$ corrections and the estimation of the number of independent samples will be combined in the following section.

\subsection{Numerical simulation and setup}

We first discuss the numerical setup before diving into the components of the numerical results. As explained earlier, when choosing $K$ parameter to perform the large $\NC$ limit, there are several requirements. In~\ref{appendix:numparam}, we give the values of $K$ that have been used.

We evaluate the perturbative coefficients of the internal energy and its variance up to $\order{g^8}$ or equivalently $\order{\lambda^4}$.
The perturbative coefficients of the odd-order terms in $g$ should be zero in perturbation theory.
We have statistically confirmed
this point within the one-sigma level in our numerical simulation. As a result, in the content that follows, we will only be concentrating on the perturbative coefficients of the even-orders of $g$.

The hyper-parameters of the NSPT simulation algorithm, the trajectory length $t$ 
and the number of trajectory time step $N_{\mathit{md}}$, are tuned 
to satisfy the large $\NC$ factorization property on the perturbative coefficients.
This means that the variance of the perturbative coefficients vanishes in the large $\NC$ limit.
The details of the large-$\NC$ factorization property on the variance are shown in~\ref{appendix:hyperparam}.
We found that the trajectory length $t=0.05$ is enough for the large $\NC$ factorization for the first four order coefficients.

In this section, we show the results with the trajectory length at $t = 0.05$ using several $N_{\mathit{md}}$'s for the zero-step size limit.
We accumulated over 2,000,000 trajectories after discarding the first 100,000 for thermalization.
To remove the autocorrelation among the samples, we sample every trajectory and bin every 100 trajectories into a bin. 
Statistics are also shown in Tabs.~\ref{tab:simulationparam}--\ref{tab:simulationparam3} of~\ref{appendix:numparam}.

\subsection{Internal energy}

In the large $\NC$ limit, the first three order results can be compared with the analytic formula for the PCM on an infinity volume lattice~\cite{weakcouplingPCM,weakcouplingPCM2}.
Because there is no
analytic formula, the fourth coefficient of the TRPCM cannot be directly compared.
To have an insight into the large $\NC$ limit of $\expval{E^{(4)}}$, 
we can employ the result from the NSPT simulation of PCM done by Bruckmann--Puhr~\cite{Bruckmann:2019mky}, 
where the raw data are available in Ref.~\citen{BruckmannBuhrNSPTdata}. 
We also perform another NSPT simulation with PCM because the Bruckmann--Puhr's data has fewer statistics to compare to our TRPCM result.
~\ref{appendix:NSPTPCM} contains a detailed description of our NSPT simulation of PCM.

\newcommand{\figscale}{0.25}
\begin{figure}[t]
\centering
\includegraphics[clip,scale=\figscale]{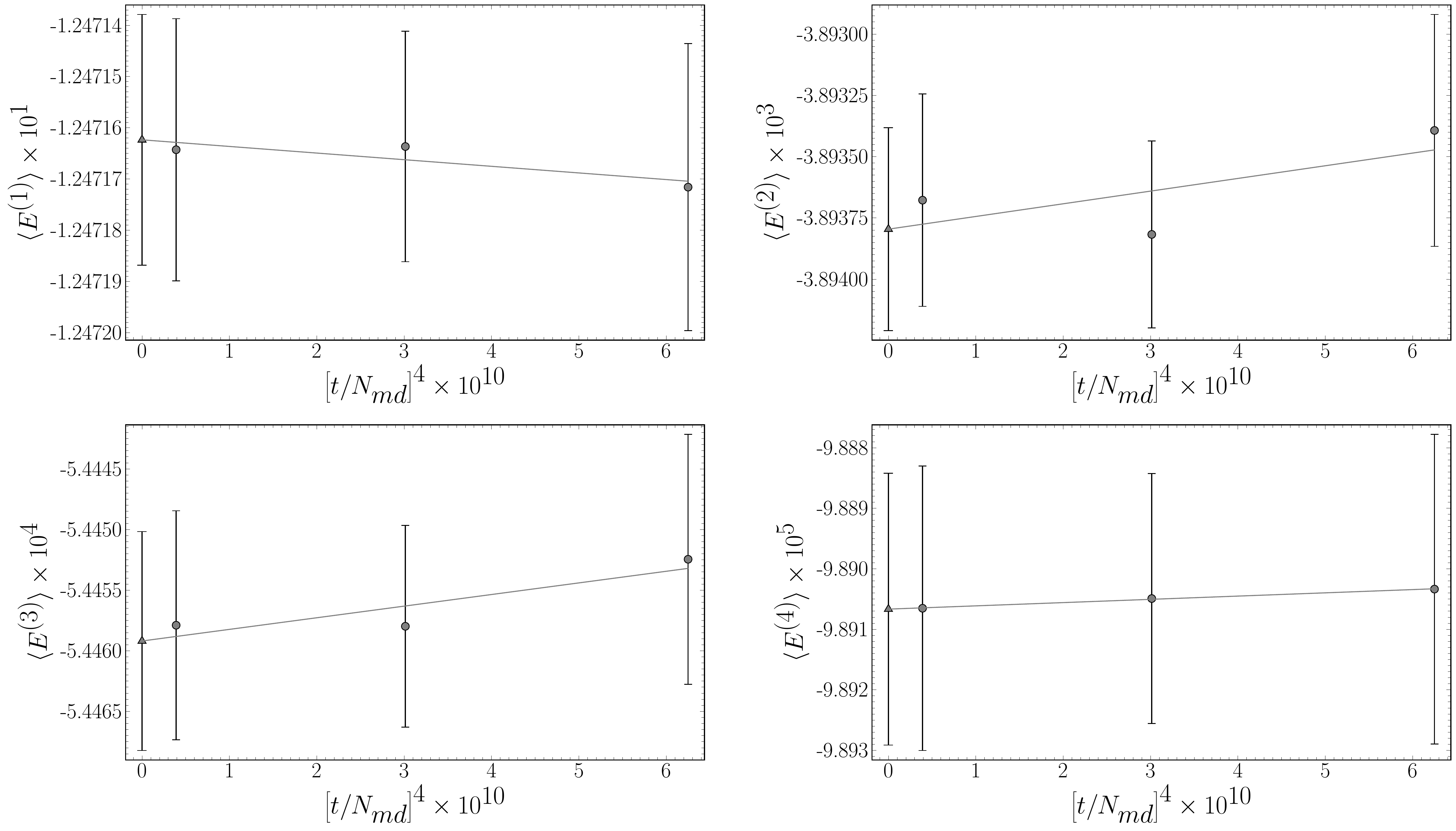}
\caption{An overview of extrapolation to vanishing MD step size ($\qty[t/N_{\mathit{md}}]^4 = \delta t^4 \to 0$). 
         These plots show our numerical results obtained from $\NC=21$ up to fourth order. 
         The perturbative coefficients at $\delta t = 0$ are extracted with linear fitting in $\qty[t/N_{\mathit{md}}]^4$.}
 \label{fig:extrapdtzero}
\end{figure}

\begin{table}[t]
\tbl{Perturbative coefficients for the internal energy after extrapolating to vanishing integration step size.}
{
\begin{tabular}{
r
S[table-format=2.12]
S[table-format=2.12]
S[table-format=2.12]
S[table-format=2.12]
} \toprule
$\NC$ & 
{$\expval{E^{(1)}}$}  & 
{$\expval{E^{(2)}}$}  & 
{$\expval{E^{(3)}}$}  & 
{$\expval{E^{(4)}}$}  \\
\colrule
 3 &    -0.11112234(1477) &-0.00328153(185)       &-0.00049017(43)  &-0.00010100(13)       \\ 
 5 &    -0.12000053(808)  &-0.00367717(110)       &-0.00052990(23)  &-0.00010180(6)        \\ 
 7 &    -0.12245450(527)  &-0.00379498(75)        &-0.00054004(16)  &-0.00010128(4)        \\ 
 9 &    -0.12345411(379)  &-0.00384602(56)        &-0.00054238(12)  &-0.00010108(3)       \\ 
11 &    -0.12396632(289)  &-0.00386240(44)        &-0.00054371(9)   &-0.00009996(2)        \\ 
13 &    -0.12425785(230)  &-0.00387216(36)        &-0.00054340(8)   &-0.00009976(2)        \\ 
15 &    -0.12444483(188)  &-0.00388285(30)        &-0.00054445(6)   &-0.00009937(2)        \\ 
17 &    -0.12456844(226)  &-0.00388773(37)        &-0.00054450(8)   &-0.00009915(2)        \\ 
19 &    -0.12465177(166)  &-0.00389026(28)        &-0.00054433(6)   &-0.00009896(1)        \\ 
21 &    -0.12471624(245)  &-0.00389380(41)        &-0.00054459(9)   &-0.00009891(2)        \\ 
\botrule
\end{tabular} \label{table:inf}}
\end{table}

Fig.~\ref{fig:extrapdtzero} is an example of the extrapolation to vanishing MD step size for the internal energy up to the fourth-order at $\NC=21$.
With our choice of the MD integrator, 4th order OMF, we linearly extrapolate with $dt^4$. 
We found similar linear dependencies for the other $\NC$ we investigated, removing the systematic error caused by the finite fictitious 
time step size. Following that, we will concentrate on the results at vanishing MD step size.
Table~\ref{table:inf} shows the perturbative coefficients of the internal energy at vanishing MD step size for each $\NC$.

\begin{figure}[t]
\centering
\includegraphics[clip,scale=\figscale,trim= 10 0 0 0]{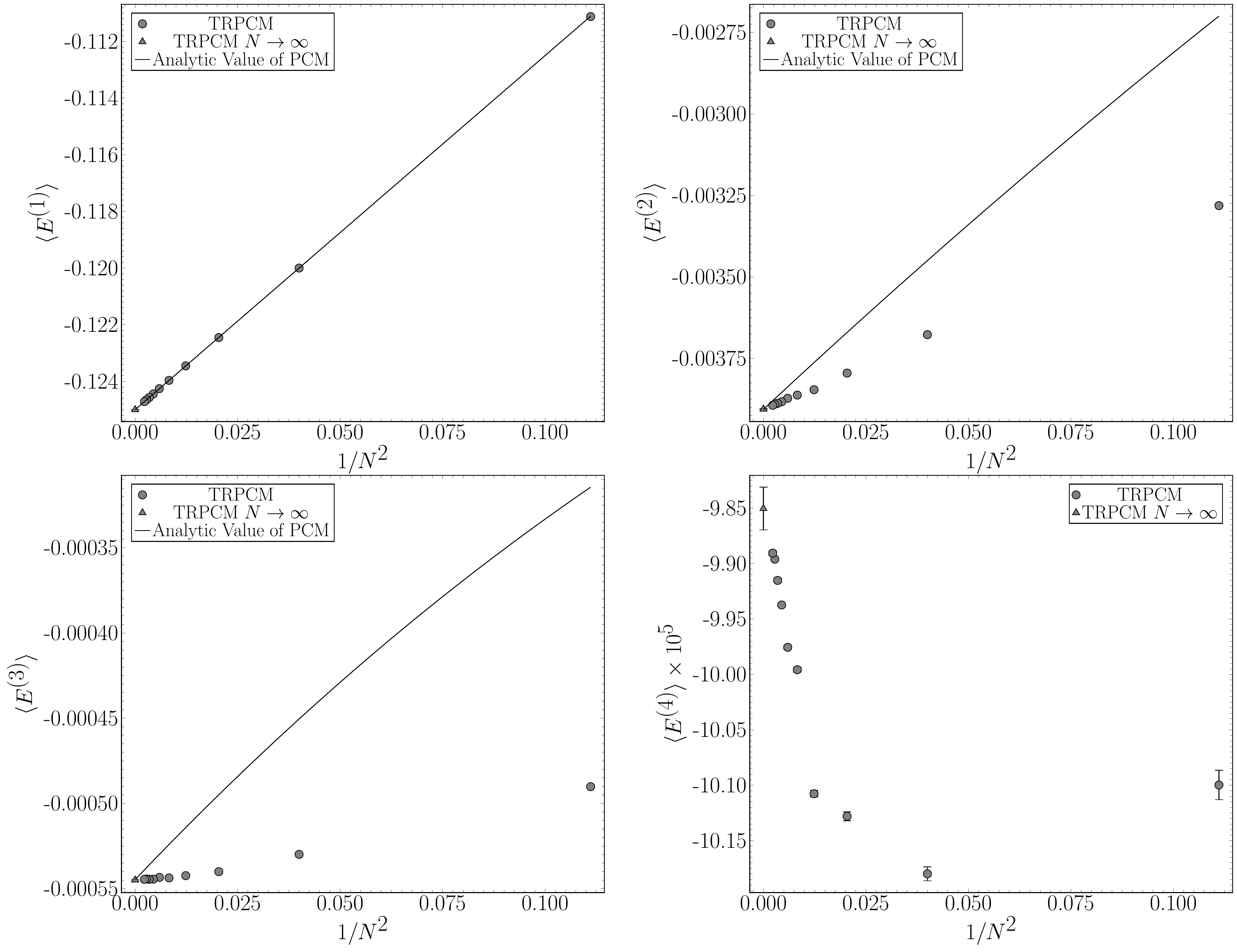}
\caption{$\NC$ dependence of perturbative coefficients of the internal energy up to fourth order. 
         The solid lines are plotted for  Eqs.~\eqref{eq:PCMstart}--\eqref{eq:PCM} as a function of $1/\NC^2$.}
 \label{fig:resfar}
\end{figure}

\begin{figure}[t]
\centering
\includegraphics[clip,scale=0.38,trim = 8 0 0 0 ]{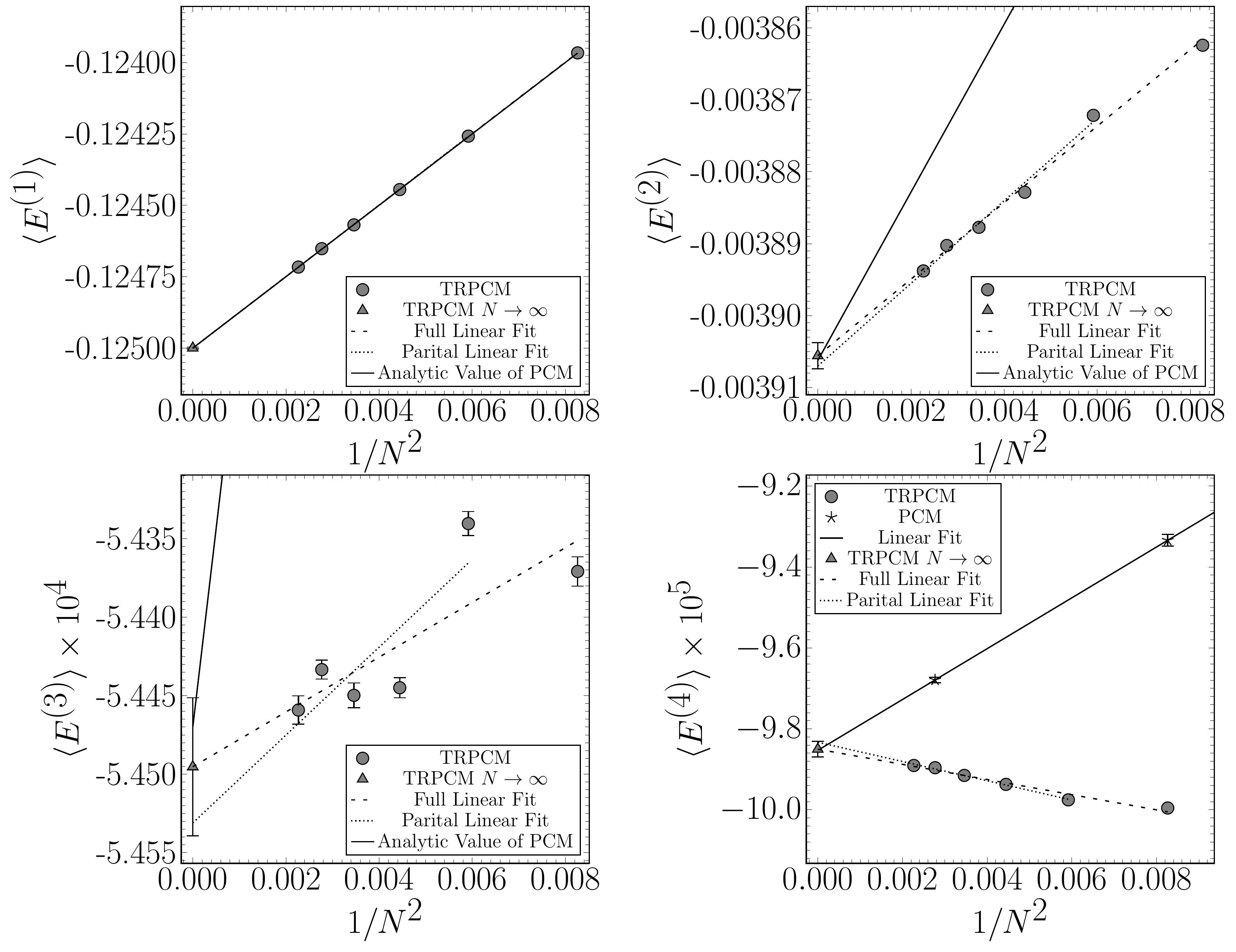}
\caption{Magnification of Fig.~\ref{fig:resfar} to show the limits as $\NC\to\infty$. }
 \label{fig:resclose}
\end{figure}

Fig.~\ref{fig:resfar} depicts the $\NC$ dependence of the perturbative coefficient of the internal energy up to the fourth-order for all $\NC$ data.
The circles represent the TRPCM with NSPT results, and the solid curves present the analytic formula\cite{weakcouplingPCM} 
for the PCM on an infinite volume lattice. 
The analytic values of the PCM coefficients up to the third order in an infinite volume are
\begin{align}
\label{eq:PCMstart}
  \expval{E^{(1)}} &= -\frac{\NC^{2}-1}{8 \NC^{2} },\\
  \expval{E^{(2)}} &= -\frac{\NC^{2}-1}{8 \NC^{2} }\times \frac{\NC^{2}-2}{32 \NC^{2}},\\
  \expval{E^{(3)}} &= -\frac{\NC^{2}-1}{8 \NC^{2} }\qty[  \frac{3 \NC^{4}-14 \NC^{2}+20}{768 \NC^{4}}+\frac{\NC^{4}-4 \NC^{2}+12}{64 \NC^{4}} Q_{1}+\frac{\NC^{4}-8 \NC^{2}+24}{64 \NC^{4}} Q_{2} ],
\label{eq:PCM}
\end{align}    
where $Q_{1}=\num{0.0958876}$ and $Q_{2}=-\num{0.0670}$\cite{weakcouplingPCM}.
In~\ref{appendix:ananl}, we also provide the analytic  values for the second order coefficient of the TRPCM for our values of $\NC$ and $K$, demonstrating 
complete consistency of the NSPT results with them.
Fig.~\ref{fig:resfar} shows that the leading order results are identical to the PCM, 
whereas the second and third-order coefficients differ from the PCM at finite $\NC$.
The $\NC$ dependence of the TRPCM is milder than that of PCM indicating that 
the large $\NC$ limit can be taken efficiently with the TRPCM.
The effectiveness of the TRPCM is more enhanced by this smaller $\NC$ dependence because 
the PCM needs double large limits on volume and $\NC$ when taking the large $\NC$ limit.

In the TRPCM, the $\NC$ dependence corresponds to  both finite $\NC$ and finite volume corrections.
We fit the NSPT data linearly in $1/\NC^2$ and the dashed and dotted lines represent the fit results as shown in Fig.~\ref{fig:resclose}. 
As we focus on the leading finite $\NC$ correction, we fit data with $\NC \ge 11$. 
Dashed (dotted) lines show the fit with $\NC \ge 11$ ($\NC \ge 13$), respectively. 
The upper triangle at $\NC\to\infty$ corresponds to the extrapolation with the dashed fit and the error bar contains 
the statistical and systematic errors. The difference between the dotted and dashed fittings is used to calculate the systematic error.
Table~\ref{table:res} shows the TRPCM's large $\NC$ limit for the first four coefficients of the internal energy.
The first three coefficients, $\expval{E^{(k)}}, k=1,2,3$, are consistent with  the analytic results of PCM in the large $\NC$ limit.

For the fourth-order coefficient, we compare the NSPT results between the PCM and TRPCM. 
We calculate the PCM coefficient at $\NC=11$ and $19$ at various volumes and use the infinite volume limit (for the details see~\ref{appendix:NSPTPCM}). 
The linear extrapolation on the PCM (solid line on stars in the bottom right panel of Fig.~\ref{fig:resclose}) is consistent with that of the TRPCM. 
We also note that the slope of the TRPCM is smaller than that of the PCM indicating the smallness of the finite $\NC$ correction of the  former.

\begin{table}[t]
\tbl{Comparison between analytic and NSPT results for $\NC \to \infty$.
     The first and second errors are the statistical error and the systematic errors from the fitting range, respectively.}
{\begin{tabular}{
c
S[table-format=2.8]
S[table-format=3.12]
c
} \toprule
{Order $i$ of $\expval{E^{(i)}}$} & {Analytic}  & {NSPT} & {\ }\\
\colrule
1 &  -0.125       & -0.12499981(234){(83)}   \\ 
2 &  -0.00390625  & -0.00390558(38){(143)}    \\ 
3 &  -0.00054470  & -0.000544953(82){(358)}   \\ 
4 &   {N.A.}      & -0.000098504(20){(172)}   \\ \botrule
\end{tabular}
\label{table:res}}
\end{table}

Beyond the leading order, the finite $\NC$ correction appears.~\cite{TEKandOG} The linear fitting results using $\NC \ge 11$ are
\begin{align}  
   \expval{E^{(1)}} &= -\num{0.12499981(248)} +\frac{\num{ 0.12510(52)}}{\NC^{2}}, 
\label{eq:oneloopform1} 
\\
   \expval{E^{(2)}} &= -\num{0.00390558(147)}  +\frac{\num{ 0.005320(82)}}{\NC^{2}},\\
   \expval{E^{(3)}} &= -\num{0.000544953(367)} +\frac{\num{ 0.000174(17)}}{\NC^{2}},\\
   \expval{E^{(4)}} &= -\num{0.000098504(173)} +\frac{\num{  -0.0001874(42)}}{\NC^{2}}.
\label{eq:oneloopform} 
\end{align}
The coefficients of $\order{1/\NC^2}$ term for $\expval{E^{(2)}}$ and $\expval{E^{(3)}}$ are  smaller than those of PCM ($3/256\simeq \num{0.01771875}$ for $\expval{E^{(2)}}$ and $\num{0.002525595}$ for $\expval{E^{(3)}}$).

We also noted that, for each order of perturbation, the magnitude of the coefficient of the $1/\NC^2$ term has the same order as that of 
the constant term.
By performing a single simulation at a finite but sufficiently large $\NC$, 
where $\NC$ is determined by requiring that the magnitude of the finite $\NC$ correction is smaller than that of the statistical error, 
we can safely evaluate the coefficients at the large $\NC$ limit.
As we will see in the following subsection, the variance and the large $\NC$ factorization can both have an impact on the relative magnitude of the statistical error with the finite $\NC$ correction. In this way, we can estimate the number of statistical samples required to get a statistical error exceeding the finite $\NC$ correction of a single large value of $\NC$. This issue will be covered in section~\ref{sec:outlook}.

\subsection{Large $\NC$ factorization and statistical error}

Using NSPT, we were able to obtain appropriate fit lines in the previous subsection that provided consistent values
for the large $\NC$ limit for the first four order coefficients. 
In order to see the confidence in the large $\NC$ limit, 
we also validate another important property of large $\NC$ field theory called 
large $\NC$ factorization\cite{factorization,factorization2}. 

According to the large $\NC$ factorization, the expectation value of the product of single trace local operators at different sites
becomes the product of each expectation value of the local operators.
The finite $\NC$ correction to the factorization scales as a function $f(1/\NC) \to 0$ in $\NC\to \infty$.
This property can be checked by observing the statistical variance of a local operator in the TRPCM 
as it corresponds to the finite $\NC$ correction as seen below.

The statistical error is proportional to the square root of the
variance and inversely proportional to the number of independent samples of the
Markov chain Monte Carlo simulation. The large N factorization
property leads to a decrease of the variance with $N$ implying also a
reduction in the number of statistical samples needed to achieve a
certain precision with NSPT for the perturbative coefficients of the
internal energy.

The factorization property indicates that the variance of the internal energy, $\mathrm{Var}\left[E\right]$, 
should behave as 
\begin{align}
   \mathrm{Var}\left[E\right] = \expval{E^2} - \expval{E}^2 
\quad& \underset{\NC \to \infty}{\rightarrow} \quad 0.
\end{align}

The same factorization property holds for each perturbation coefficient $\expval{E^{(k)}}$.
As one of the validations of the outcomes assessed with NSPT, we use this property to examine the variance of the internal energy in each order 
by fitting the $1/\NC^2$ dependence.

Fig.~\ref{fig:var} and Table~\ref{table:var}  depict the $\NC$ dependence of the variance.
Fig.~\ref{fig:var2} is the magnification of Fig.~\ref{fig:var} for the large $\NC$ limit.
We fit the variance as a linear function of $1/\NC^2$ as
\begin{align}
 \mathrm{Var}(E^{(k)}) = \mathrm{Var}(E^{(k)})_{\NC\to\infty} +  \dfrac{a^{(k)}}{\NC^2}.
\end{align}
The dashed and dotted lines are fit results with and without $\NC=11$ data, respectively.
We clearly observe the vanishing variance in the $N \to \infty$ limit.
The up-triangles at $1/\NC^2 = 0$ are the results without $\NC=11$ data and the error includes statistical 
and systematic errors assigned with the discrepancy between the two fit results.
Table~\ref{table:varres} shows the fit results.
The fitting results for $\mathrm{Var}(E^{(k)})_{\NC\to\infty}$ are consistent with zero 
where the first and the second errors are the statistical and the systematic errors, respectively. This shows that our simulation setup effectively maintains the large $\NC$ factorization property.
In light of this, these results provide a precise cross-check for the reliability of our NSPT results up to the fourth-order. Additionally,  we see that as order $k$ increases, the slope $a^{(k)}$, where we only mention the statistical error, decreases.

\renewcommand{\figscale}{0.27}
\begin{figure}[t]
    \centering  
    \includegraphics[clip,scale=\figscale,trim= 8 0 0 0]{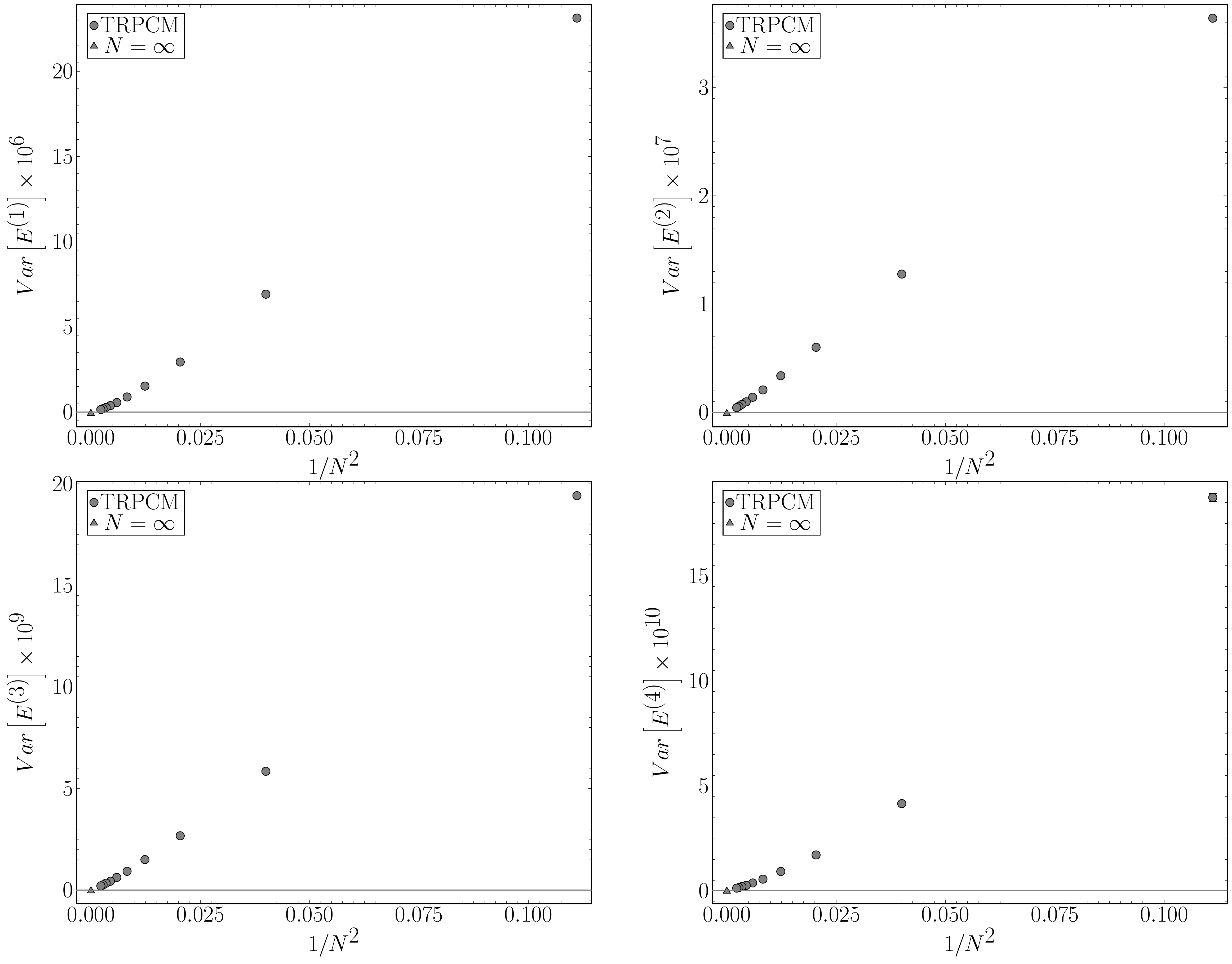}
    \caption{The $\NC$ dependence of variance of perturbative coefficients 
             of the internal energy up to fourth order.}
    \label{fig:var}
\end{figure}

\renewcommand{\figscale}{0.36}
\begin{figure}[t]  
    \centering  
    \includegraphics[clip,scale=\figscale,trim= 8 0 0 0]{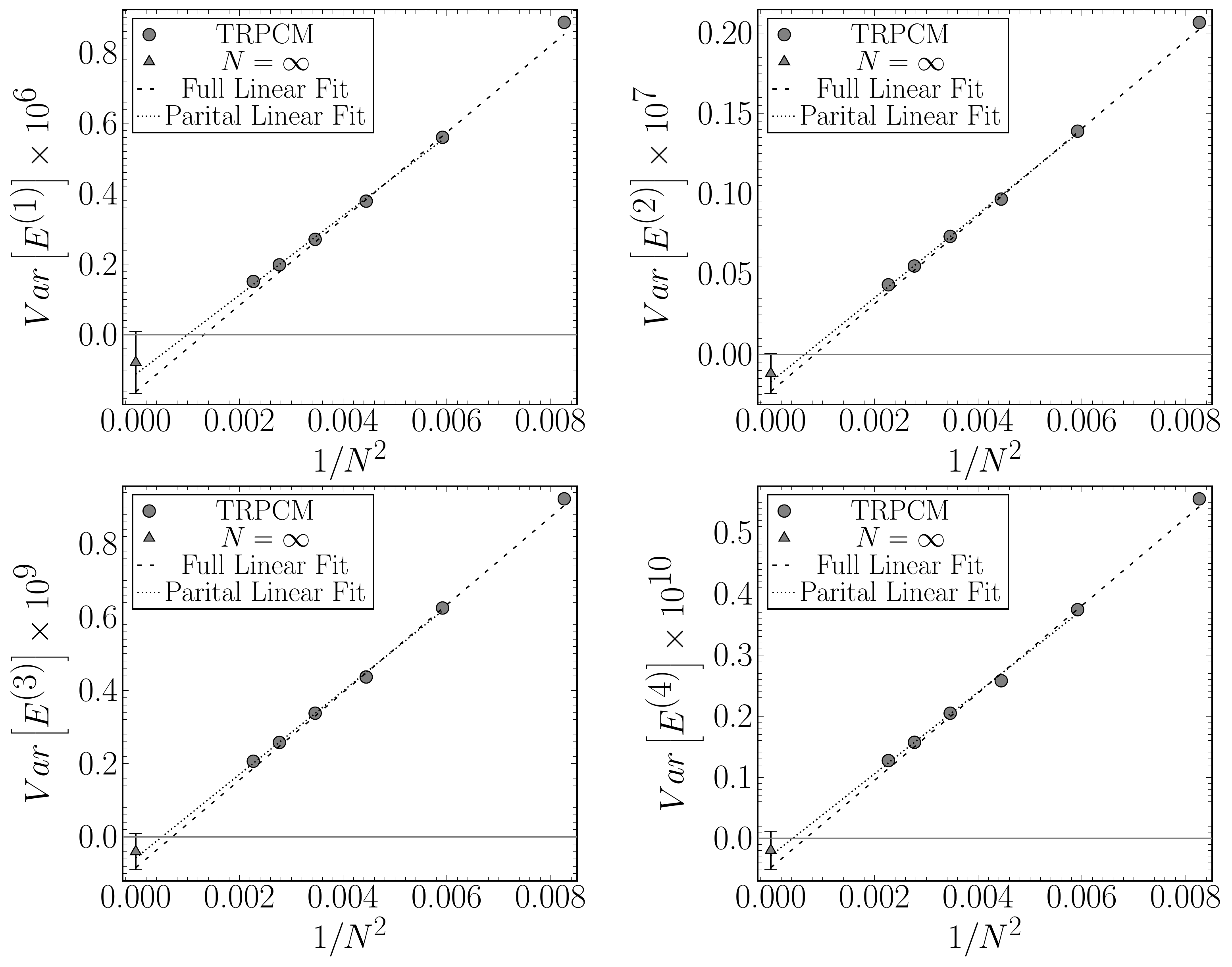}
    \caption{Magnification of Fig.~\ref{fig:var} to show the limits as $\NC\to\infty$.}
    \label{fig:var2}
\end{figure}

\begin{table}[t]
\tbl{Variance of perturbative coefficients of the internal energy after 
     extrapolating to vanishing integration step size.}
{\begin{tabular}{
r
S[table-format=1.12]
S[table-format=1.14]
S[table-format=1.14]
S[table-format=1.14]
} \toprule
{$\NC$} 
& {$\mathrm{Var}[E^{(1)}]\times 10^4$}  
& {$\mathrm{Var}[E^{(2)}]\times 10^6$}
& {$\mathrm{Var}[E^{(3)}]\times 10^7$}
& {$\mathrm{Var}[E^{(4)}]\times 10^8$} \\
\colrule
 3 &    0.2312(10)   &0.3639(17)    &0.1941(12)    &0.1874(19)   \\
 5 &    0.06919(30)  &0.12763(56)   &0.05845(27)   &0.04153(22)  \\
 7 &    0.02938(12)  &0.06003(26)   &0.02670(11)   &0.017085(79) \\
 9 &    0.015236(66) &0.03374(14)   &0.014971(65)  &0.009203(41) \\
11 &    0.008863(38) &0.020659(90)  &0.009228(40)  &0.005554(24) \\
13 &    0.005601(24) &0.013890(61)  &0.006247(27)  &0.003741(17) \\
15 &    0.003789(16) &0.009662(42)  &0.004364(18)  &0.002578(11) \\
17 &    0.002705(14) &0.007339(39)  &0.003375(18)  &0.002047(11) \\
19 &    0.001981(95) &0.005496(26)  &0.002575(12)  &0.0015712(75)\\
21 &    0.001510(12) &0.004325(33)  &0.002064(15)  &0.0012717(99)\\  \botrule
\end{tabular} \label{table:var}}
\end{table}

\begin{table}[t]
\tbl{Variance of internal energy after performing the extrapolation to infinite $\NC$.}
{
\begin{tabular}{cll}\toprule
{Order $i$ of $\mathrm{Var}[E^{(i)}]$} & {$\mathrm{Var}[E^{(i)}]_{\NC\to\infty}$}  & {$a^{(i)}$}\\
\colrule
1 &  -0.786(45)(834)$\times 10^{-7}$  & 0.11242(68)$\times 10^{-3}$  \\
2 &  -0.119(11)(111)$\times 10^{-8}$  & 0.2612(17)$\times 10^{-5}$   \\
3 &  -0.403(54)(440)$\times 10^{-10}$ & 0.11403(78)$\times 10^{-6}$  \\
4 &  -0.198(33)(282)$\times 10^{-11}$ & 0.6693(47)$\times 10^{-8}$   \\
\botrule
\end{tabular}
\label{table:varres}}
\end{table}

The statistical error is related to the variance and the number of statistical sample as
\begin{align}
 \qty(\delta E^{(k)})^2 = \dfrac{\mathrm{Var}(E^{(k)})}{N_{\mathrm{sample}}},
\end{align}
where $\delta E^{(k)}$ is the statistical error of $\expval{E^{(k)}}$.

The number of the statistical samples needed for a fixed relative statistical error can be estimated as
\begin{align}
N_{\mathrm{sample}} = \frac{\mathrm{Var}(E^{(k)})}{\expval{E^{(k)}}^{2}}\qty(\dfrac{\expval{E^{(k)}}}{\delta E^{(k)}})^{2}.
\label{eq:Nsample}
\end{align}
In the large $\NC$ limit, $\mathrm{Var}(E^{(k)})$ and $\expval{E^{(k)}}$
behave as
\begin{align}
 \mathrm{Var}(E^{(k)}) \simeq \dfrac{a^{(k)}}{\NC^2},
\quad
 \expval{E^{(k)}} \simeq \expval{E^{(k)}}_{\infty} + \dfrac{b^{(k)}}{\NC^2}.
\end{align}
For a fixed relative statistical error, the number of statistical samples 
scales as
\begin{align}
N_{\mathrm{sample}} \simeq 
\dfrac{1}{\NC^2}
\dfrac{a^{(k)}}{\expval{E^{(k)}}_{\infty}^2}
\qty(\dfrac{\expval{E^{(k)}}}{\delta E^{(k)}})^{2},
\label{eq:Nsample2}
\end{align}
in the large $\NC$ limit. 
For a fixed relative error, the number of statistical samples decreases as $1/\NC^2$ for the large $\NC$ limit.
This is just the so-called master field property of the large $\NC$ limit~\cite{Witten:1979pi}.
Conversely, the relative statistical error decreases as increasing $\NC$ linearly at a fixed number of samples as
\begin{align}
 \dfrac{\delta E^{(k)}}{\abs{\expval{E^{(k)}}}} = \dfrac{\sqrt{\mathrm{Var}(E^{(k)})}}{\sqrt{N_{\mathrm{sample}}}}\simeq
 \dfrac{\sqrt{a^{(k)}}}{\NC \sqrt{N_{\mathrm{sample}}}}.
 \label{eq:StatNdep}
\end{align}

In the previous subsection, we investigated the finite $\NC$ correction for $\expval{E^{(k)}}$.
The size of $\NC$ with which the finite $\NC$ correction is sufficiently smaller than 
the statistical error can be estimated. In order to estimate the number of statistics
with a single simulation for the large $\NC$ limit, we will combine the property of the finite $\NC$ correction
and the $\NC$ dependence of the relative statistical error in the following section.

\section{Outlook}
\label{sec:outlook}

The large $\NC$ results of the TRPCM will be performed in a single simulation as our ultimate objective. With a fixed number of statistical samples, as shown in Eq.~\eqref{eq:StatNdep}, the relative statistical error decreases toward the large $\NC$ limit. 
When the magnitude of the finite $\NC$ correction term $|b^{(k)}|/\NC^2$ is comparable to the statistical error of the coefficient $\expval{E^{(k)}}_{\infty}$, 
we can recognize the value of $\expval{E^{(k)}}$ at the finite $\NC$ as the value at the large $\NC$ limit. 
This means that we could obtain the large $\NC$ result of the TRPCM with a single simulation without large $\NC$ extrapolation if the statistical error and
the $\NC$ correction occur in the same order at a sufficiently large but finite $\NC$. We explore this
possibility as our outlook for the high-order NSPT calculation of the TRPCM at the large $\NC$ limit.

To do this, we have to estimate the $k$ dependence of the finite $\NC$ corrections, $a^{(k)}$ and $b^{(k)}$, larger than $4$. 
We fit the ratio $-\sqrt{a^{(k)}}/\expval{E^{(k)}}_{\infty}$ as a function of order $k$ using the data at $k>1$. 
We show our fitting results in Fig.~\ref{fig:orderdepadivEsqr}. It shows a linear behavior as 
\begin{align}
    \dfrac{\sqrt{a^{(k)}}}{\abs{\expval{E^{(k)}}_{\infty}}} = 0.207k.
    \label{eq:kdepA}
\end{align}
For the finite $\NC$ correction term, $\abs{b^{(k)}/(\NC^2\expval{E^{(k)}})}$, our observation 
based on Eqs.~\eqref{eq:oneloopform1}--\eqref{eq:oneloopform} suggests $|b^{(k)}|\sim |\expval{E^{(k)}}|$ so that the finite $\NC$ correction behaves as $1/\NC^2$. 
The Feynman diagrammatic argument, however, suggests the finite $\NC$ correction may increase as $k$ rises. In light of this, we, now, consider two possible functional forms for the finite $\NC$ correction.

\begin{align}
    \abs{\dfrac{b^{(k)}}{\expval{E^{(k)}}_{\infty}}}=1\quad\mbox{or}
  \quad
    \abs{\dfrac{b^{(k)}}{\expval{E^{(k)}}_{\infty}}}=k.
      \label{eq:kdepB}
\end{align}

\renewcommand{\figscale}{0.7}
\begin{figure}[t]  
    \centering  
    \includegraphics[clip,scale=\figscale]{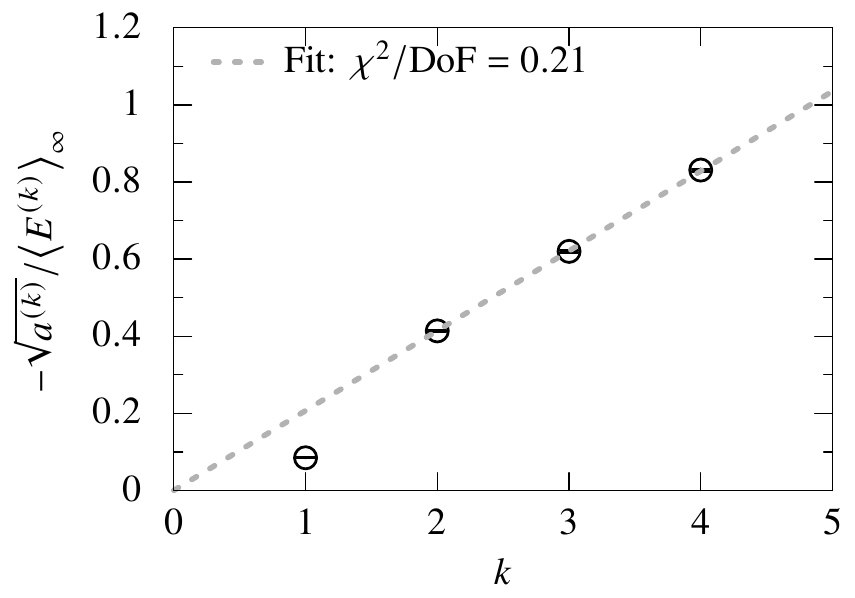}
    \caption{The $k$ dependence of $-\sqrt{a^{(k)}}/\expval{E^{(k)}}_{\infty}$.}
    \label{fig:orderdepadivEsqr}
\end{figure}

Now we analyze the feasibility of a single simulation at a large but finite $\NC$ 
for extracting the perturbation coefficients in the large $\NC$ limit at a higher order.
We set the desired order to be $k=20$, which order enables us to observe the renormalon behavior as in Ref.~\citen{Bruckmann:2019mky}, 
and the relative error as $\delta E^{(k=20)}/\abs{\expval{E^{(k=20)}}} = 1\%$.
We also fix the size of $\NC$ to be 100 for example.

\begin{figure}[t]  
    \centering  
    \includegraphics[clip,scale=\figscale]{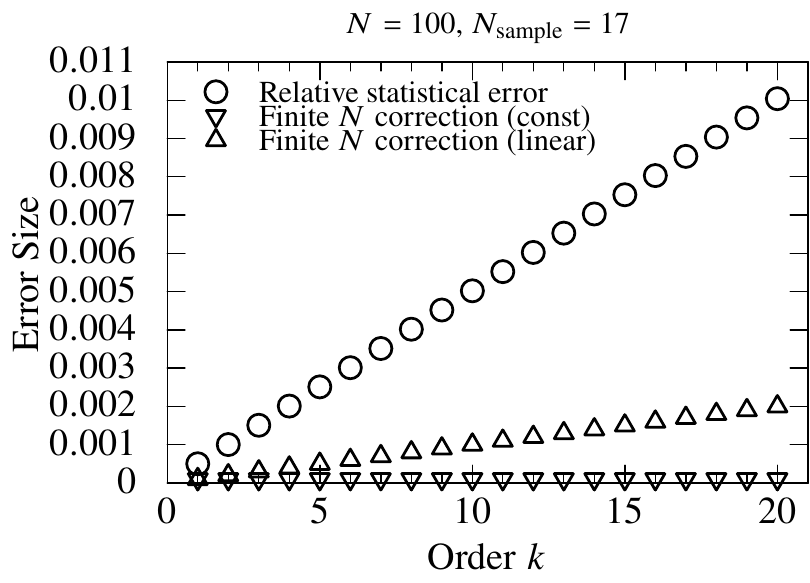}
    \includegraphics[clip,scale=\figscale]{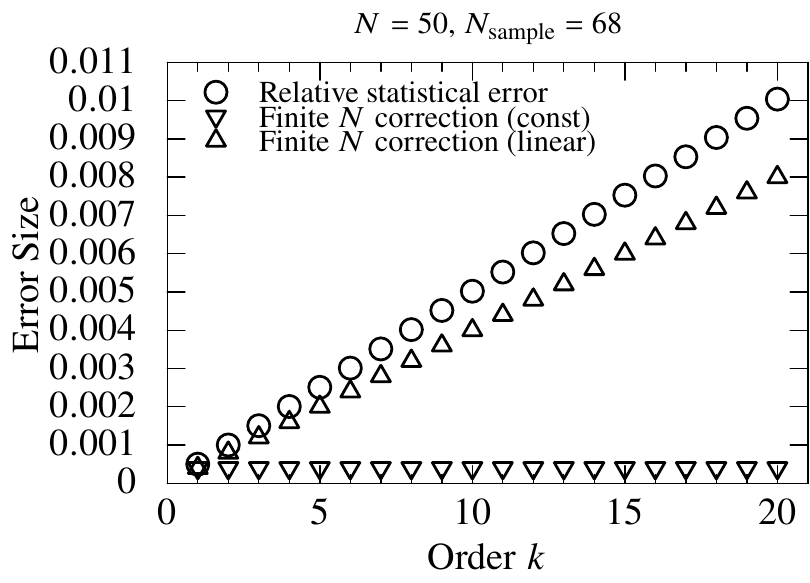}
    \caption{The $k$ dependence of relative statistical error and finite $\NC$ correction. 
             The number of independent samples is determined by fixing 
            the relative statistical error to be $1\%$ at order $k=20$.}
    \label{fig:orderdepErrors}
\end{figure}

Using Eq.~\eqref{eq:Nsample2}, we obtain
the number of samples needed to reach a relative statistical error for an order $k$ 
as 
\begin{align}
    N_{\mathrm{sample}} \simeq 
\left(\dfrac{0.207k}{\NC}\right)^2
\qty(\dfrac{\expval{E^{(k)}}}{\delta E^{(k)}})^{2}.
\end{align}

Substituting the settings, $k=20$ and $\delta E^{(k=20)}/\abs{\expval{E^{(k=20)}}} = 1\%$ into this equation,
we obtain $N_{\mathrm{sample}}  = 17$. 
With $N_{\mathrm{sample}}  = 17$, we investigate the possibility that the finite $\NC$ correction is comparable to or smaller than the statistical error at each order lower than $k=20$. 
The left panel in Fig.~\ref{fig:orderdepErrors} shows the $k$ dependence of the relative statistical error and finite $\NC$ correction with $\NC=100$.
The circles are the relative statistical error with $N_{\mathrm{sample}}  = 17$, which is determined as 1\% at $k=20$. 
The up and low triangles are the finite $\NC$ correction with constant and linear assumptions \eqref{eq:kdepB}, respectively. Both of the finite $\NC$ corrections are sufficiently smaller than the statistical error so that the single simulation at  $\NC=100$ with $N_{\mathrm{sample}}  = 17$ effectively yields the perturbative coefficients up to the 20th order in the large $\NC$ limit.

On the other hand, the finite $\NC$ correction increases for smaller $\NC$, making it statistically visible. The right panel in Fig.~\ref{fig:orderdepErrors} shows the $k$ dependence of the relative statistical error and finite $\NC$ correction with $\NC=50$, 
where $N_{\mathrm{sample}}  = 68$ is determined in the same manner to $\NC=100$.
The finite $\NC$ correction with linear dependence approaches the statistical error. 
Below $\NC=50$, the finite $\NC$ correction could be statistically visible with the linear dependence assumption.

As was mentioned above, setting the relative statistical error at a higher order coefficient and making the assumption that the finite $\NC$ corrections depend on $k$,
the large $\NC$ limit could be reached with a single NSPT simulation at a large enough $\NC$ value.

\section{Summary}
In this study, we have applied NSPT to the TRPCM and evaluated the perturbative coefficients of the internal energy up to $\mathcal{O}(g^8)$ or equivalently $\mathcal{O}(\lambda^4)$. 
We have shown that at the large $\NC$ limit the first three order coefficients agree exactly with the analytic results of the
lattice PCM on an infinite volume lattice. 
With the help of an independent NSPT simulation of the lattice PCM, we were able to accurately extract the fourth-order coefficient in the large $\NC$ limit.
We discussed the possibility of $N=\infty$ limit simulation with a single NSPT simulation at a sufficiently large but finite $\NC$ value from the consistent large $\NC$ factorization property of the observable.
We investigated whether the $N=\infty$ simulation could be successfully carried out $\NC=100$
with a 1\% relative statistical error at the 20th order coefficient, 
where the finite $\NC$ corrections could be statistically invisible under the assumption that they depend only on the order of the perturbation, rather than having a more complex dependence. To demonstrate the benefit of the TRPCM that the large $\NC$ and large volume limits are taken effectively with a single simulation, more NSPT simulation at larger $\NC$ and higher order should be run.

\section{Acknowledgments}
{
We thank Falk Bruckmann for pointing us the location of the raw data of their PCM work.
A.G.-A. is partially supported by grant PGC2018-094857-B-I00 funded by MCIN/AEI/ 10.13039/501100011033 and by ``ERDF A way of making Europe'', 
and by the Spanish Research Agency (Agencia Estatal de Investigaci\'{o}n) through grants IFT Centro de Excelencia Severo Ochoa SEV-2016-0597 and No CEX2020-001007-S,
funded by MCIN/AEI/10.13039/501100011033. 
He also acknowledges support from the project H2020-MSCAITN-2018-813942 (EuroPLEx) and the EU Horizon 2020 research and innovation programme, 
STRONG-2020 project, under grant agreement No 824093.
K.-I.I. is supported by MEXT as ``Program for Promoting Researches on the Supercomputer Fugaku''
(Simulation for basic science: from fundamental laws of particles to creation of nuclei, JPMXP1020200105) and JICFuS.
M.O. is supported by JSPS KAKENHI Grant Number 21K03576.
The computation was carried out using the computer resource offered under the category of General Projects 
by Research Institute for Information Technology, Kyushu University.
}

%\newpage
\appendix
%\appendixpage
%\addappheadtotoc
%\renewcommand\chaptername{Appendix}
%\begin{appendices}

%%%%%%%%% LONG TABLE FOR THE HYPER PARMETER AND STATISTICS
\begin{table}[t]
\tbl{Simulation parameters and statistics for $N=3$--$13$}
{ \begin{tabular}{SSSSSSl}
\toprule
        {$\NC$}                    &
        {$\qty(\dfrac{L}{a})^2$}   & 
        {$K$}                      & 
        {$\bar{K}$}                & 
        {$\dfrac{\bar{K}}{\NC}$}   &
        {$t$}                      &
        {($N_\mathit{md}$, \textbf{Statistics})} \\ \midrule
{\multirow{1}{*}{3}}   & {\multirow{1}{*}{9}}      & {\multirow{1}{*}{1}}    & {\multirow{1}{*}{1}}    & {\multirow{1}{*}{0.33}}    & {\multirow{1}{*}{0.05}}       & (10, 10 000 000) (12, 10 000 000) (20, 10 000 000)  \\ \midrule
{\multirow{1}{*}{5}}   & {\multirow{1}{*}{25}}     & {\multirow{1}{*}{3}}    & {\multirow{1}{*}{2}}    & {\multirow{1}{*}{0.4}}     & {\multirow{1}{*}{0.05}}       & (10, 10 000 000) (12, 10 000 000) (20, 10 000 000)  \\ \midrule
{\multirow{1}{*}{7}}   & {\multirow{1}{*}{49}}     & {\multirow{1}{*}{5}}    & {\multirow{1}{*}{3}}    & {\multirow{1}{*}{0.43}}    & {\multirow{1}{*}{0.05}}       & (10, 10 000 000) (12, 10 000 000) (20, 10 000 000)  \\ \midrule
{\multirow{1}{*}{9}}   & {\multirow{1}{*}{81}}     & {\multirow{1}{*}{7}}    & {\multirow{1}{*}{4}}    & {\multirow{1}{*}{0.44}}    & {\multirow{1}{*}{0.05}}       & (10, 10 000 000) (12, 10 000 000) (20, 10 000 000)  \\ \midrule
{\multirow{6}{*}{11}}  & {\multirow{6}{*}{121}}    & {\multirow{6}{*}{3}}    & {\multirow{6}{*}{4}}    & {\multirow{6}{*}{0.36}}    & {\multirow{1}{*}{0.025}}      & (5,   1 000 000) (6,   1 000 000) (10,  1 000 000)  \\ \cmidrule{6-7}
                       &                           &                         &                         &                            & {\multirow{1}{*}{0.05}}       & (10, 10 000 000) (12, 10 000 000) (20, 10 000 000)  \\ \cmidrule{6-7}
                       &                           &                         &                         &                            & {\multirow{1}{*}{0.08}}       & (16,  1 000 000) (19,  1 000 000) (32,  1 000 000)  \\ \cmidrule{6-7}
                       &                           &                         &                         &                            & {\multirow{1}{*}{0.1}}        & (20,  1 000 000) (24,  1 000 000) (40,  1 000 000)  \\ \cmidrule{6-7}
                       &                           &                         &                         &                            & {\multirow{1}{*}{0.2}}        & (40,    500 000) (48,    500 000) (80,    500 000)  \\ \cmidrule{6-7}
                       &                           &                         &                         &                            & {\multirow{1}{*}{0.5}}        & (100,   500 000) (120,   500 000) (200,   500 000)  \\ \midrule
{\multirow{6}{*}{13}}  & {\multirow{6}{*}{169}}    & {\multirow{6}{*}{8}}    & {\multirow{6}{*}{5}}    & {\multirow{6}{*}{0.38}}    & {\multirow{1}{*}{0.025}}      & (5,   1 000 000) (6,   1 000 000) (10,  1 000 000)  \\ \cmidrule{6-7}
                       &                           &                         &                         &                            & {\multirow{1}{*}{0.05}}       & (10, 10 000 000) (12, 10 000 000) (20, 10 000 000)  \\ \cmidrule{6-7}
                       &                           &                         &                         &                            & {\multirow{1}{*}{0.08}}       & (16,  1 000 000) (19,  1 000 000) (32,  1 000 000)  \\ \cmidrule{6-7}
                       &                           &                         &                         &                            & {\multirow{1}{*}{0.1}}        & (20,  1 000 000) (24,  1 000 000) (40,    900 000)  \\ \cmidrule{6-7}
                       &                           &                         &                         &                            & {\multirow{1}{*}{0.2}}        & (40,    500 000) (48,    500 000) (80,    500 000)  \\ \cmidrule{6-7}
                       &                           &                         &                         &                            & {\multirow{1}{*}{0.5}}        & (100,   800 000) (120,   600 000) (200,   500 000)  \\
\bottomrule
\end{tabular}}
\label{tab:simulationparam}
\end{table}

\begin{table}[t]
\tbl{Same as table~\ref{tab:simulationparam}, but for for $N=15$--$21$}
{ \begin{tabular}{SSSSSSl}
\toprule
        {$\NC$}                    &
        {$\qty(\dfrac{L}{a})^2$}   & 
        {$K$}                      & 
        {$\bar{K}$}                & 
        {$\dfrac{\bar{K}}{\NC}$}   &
        {$t$}                      &
        {($N_\mathit{md}$, \textbf{Statistics})} \\ \midrule
{\multirow{6}{*}{15}}  & {\multirow{6}{*}{225}}    & {\multirow{6}{*}{4}}    & {\multirow{6}{*}{4}}    & {\multirow{6}{*}{0.27}}    & {\multirow{1}{*}{0.025}}      & (5,   1 000 000) (6,   1 000 000) (10,  1 000 000)  \\ \cmidrule{6-7}
                       &                           &                         &                         &                            & {\multirow{1}{*}{0.05}}       & (10, 10 000 000) (12, 11 000 000) (20, 10 000 000)  \\ \cmidrule{6-7}
                       &                           &                         &                         &                            & {\multirow{1}{*}{0.08}}       & (16,  1 000 000) (19,  1 000 000) (32,  1 000 000)  \\ \cmidrule{6-7}
                       &                           &                         &                         &                            & {\multirow{1}{*}{0.1}}        & (20,  6 000 000) (24,  3 000 000) (40,  3 000 000)  \\ \cmidrule{6-7}
                       &                           &                         &                         &                            & {\multirow{1}{*}{0.2}}        & (40,    500 000) (48,    500 000) (80,    500 000)  \\ \cmidrule{6-7}
                       &                           &                         &                         &                            & {\multirow{1}{*}{0.5}}        & (100,   600 000) (120,   550 000) (200,   500 000)  \\ \midrule
{\multirow{6}{*}{17}}  & {\multirow{6}{*}{289}}    & {\multirow{6}{*}{5}}    & {\multirow{6}{*}{7}}    & {\multirow{6}{*}{0.41}}    & {\multirow{1}{*}{0.025}}      & (5,   1 000 000) (6,   1 000 000) (10,  1 000 000)  \\ \cmidrule{6-7}
                       &                           &                         &                         &                            & {\multirow{1}{*}{0.05}}       & (10,  7 240 200) (12,  7 944 400) (20,  4 684 000)  \\ \cmidrule{6-7}
                       &                           &                         &                         &                            & {\multirow{1}{*}{0.08}}       & (16,  1 000 000) (19,  1 000 000) (32,  1 000 000)  \\ \cmidrule{6-7}
                       &                           &                         &                         &                            & {\multirow{1}{*}{0.1}}        & (20,  5 000 000) (24,  5 000 000) (40,  4 000 000)  \\ \cmidrule{6-7}
                       &                           &                         &                         &                            & {\multirow{1}{*}{0.2}}        & (40,    500 000) (48,    500 000) (80,    500 000)  \\ \cmidrule{6-7}
                       &                           &                         &                         &                            & {\multirow{1}{*}{0.5}}        & (100,   800 000) (120,   800 000) (200,   600 000)  \\ \midrule
{\multirow{6}{*}{19}}  & {\multirow{6}{*}{361}}    & {\multirow{6}{*}{11}}   & {\multirow{6}{*}{7}}    & {\multirow{6}{*}{0.37}}    & {\multirow{1}{*}{0.025}}      & (5,   1 000 000) (6,   1 000 000) (10,  1 000 000)  \\ \cmidrule{6-7}
                       &                           &                         &                         &                            & {\multirow{1}{*}{0.05}}       & (10,  9 000 000) (12,  8 435 100) (20,  6 435 800)  \\ \cmidrule{6-7}
                       &                           &                         &                         &                            & {\multirow{1}{*}{0.08}}       & (16,  1 000 000) (19,  1 000 000) (32,  1 000 000)  \\ \cmidrule{6-7}
                       &                           &                         &                         &                            & {\multirow{1}{*}{0.1}}        & (20,  1 900 000) (24,  2 200 000) (40,  1 100 000)  \\ \cmidrule{6-7}
                       &                           &                         &                         &                            & {\multirow{1}{*}{0.2}}        & (40,    800 000) (48,    500 000) (80,    500 000)  \\ \cmidrule{6-7}
                       &                           &                         &                         &                            & {\multirow{1}{*}{0.5}}        & (100, 1 010 000) (120, 1 010 000) (200,   550 000)  \\ \midrule
{\multirow{1}{*}{21}}  & {\multirow{1}{*}{441}}    & {\multirow{1}{*}{8}}    & {\multirow{1}{*}{8}}    & {\multirow{1}{*}{0.38}}    & {\multirow{1}{*}{0.05}}       & (10,  2 036 200) (12,  3 084 400) (20,  2 411 300)  \\ 
\bottomrule
\end{tabular}}
\label{tab:simulationparam3}
\end{table}
%%%%%%%%%%%%%%%%%%%%%%%

\section{Numerical parameters}
\label{appendix:numparam}

This appendix contains a detailed description of the specifics of the numerical parameters. 
It also includes the NSPT parameters for the models that we are studying, as shown in Tables~\ref{tab:simulationparam}--\ref{tab:simulationparam3}. 
To take the smooth large $\NC$ limit of the TRPCM, we keep $\frac{\bar{K}}{\NC}$ about a constant which is $0.38$. 
For the NSPT, we show various trajectory lengths $t$ and $N_{\mathit{md}}$. 
Furthermore, we show the number of statistical samples. 
The number of independent samples can be obtained from the following equation
\begin{align}
N_{independent} = \frac{N_{Statistics}}{N_{bin}},
\end{align}
where, in our case, $N_{bin}$ is assigned as $100$.

%\newpage

\section{Hyper-parameters}
\label{appendix:hyperparam}

This appendix contains a description of the analysis of the NSPT simulation's hyper-parameters. The length of trajectory ($t$) and the number of MD steps ($N_{\mathit{md}}$) for the trajectory of the MD evolution serve as the hyper-parameters in this study. 
As described in the main text, the large $\NC$ factorization property is a metric for a suitable choice of the hyper-parameters.
We note that the HMD algorithm for NSPT could possess non-ergodicity as discussed in Refs.~\citen{Mackenze1989,2001,4th,4th2}.
To ensure the ergodicity, we investigate the large $\NC$ factorization property using the vanishing variance in the large $\NC$ limit.

We investigate the $t$ dependence of the variance of perturbative coefficients
evaluated at $t=0.5, 0.2, 0.1, 0.08, 0.05, 0.025$ for $11 \le \NC \le 19$.
The finite time step size error is removed by extrapolating $\delta t \to 0$
using data at three $N_{\mathit{md}}$'s at each $t$ as shown in table~\ref{tab:simulationparam}.

\renewcommand{\figscale}{0.25}
\begin{figure}[t]  
    \centering  
    \includegraphics[clip,scale=\figscale,trim= 8 0 0 0]{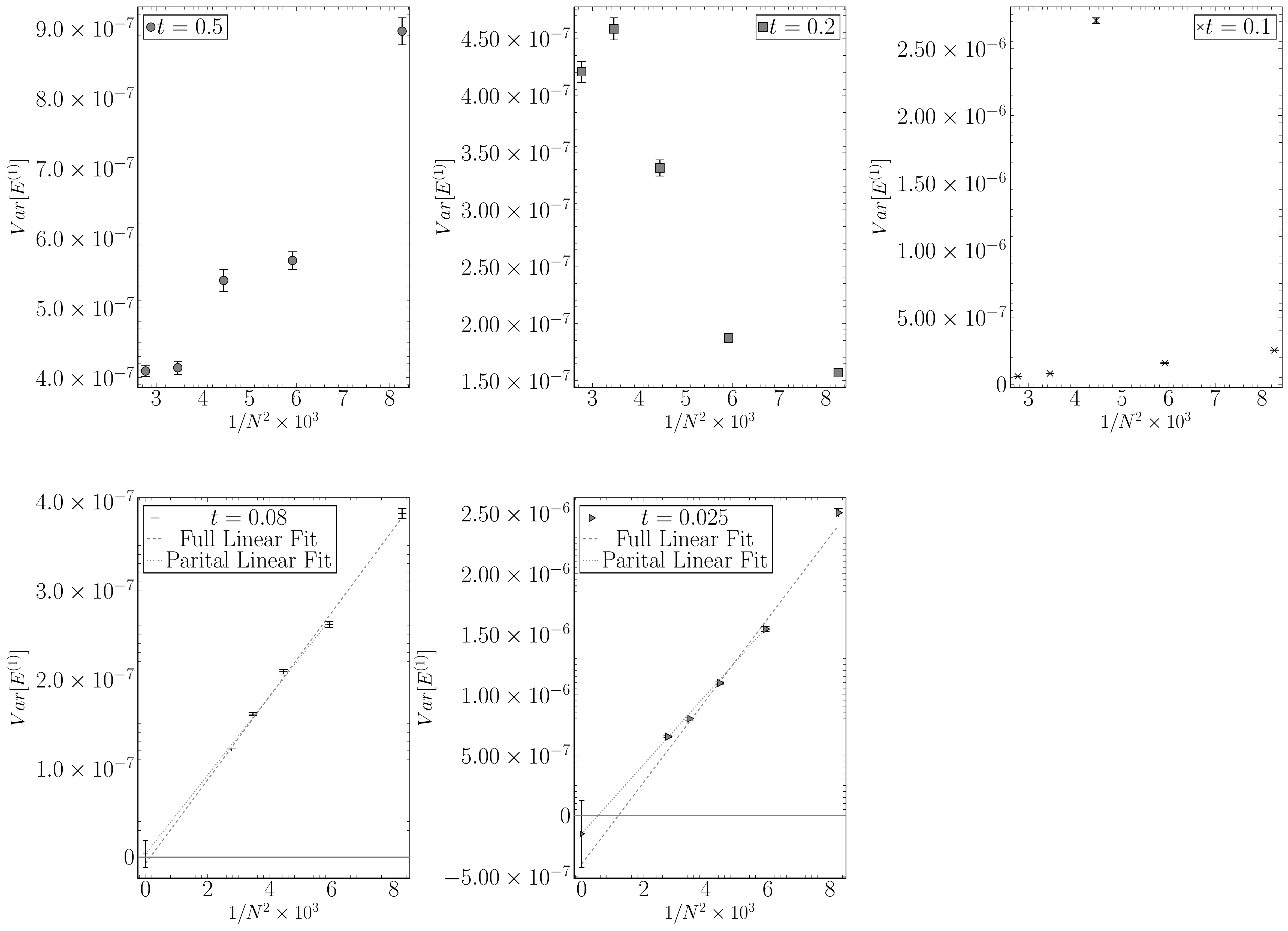}
    \caption{The hyper-parameter dependency of variance of internal energy at the leading order case. 
             For $t=0.08$ and $0.025$ data,  linear fit in $1/\NC^2$ is applied.
             The errors as $\NC\to\infty$ include the statistical error and systematic error assigned 
             by fittings with and with out of the data at $\NC=11$.}
    \label{fig:hyper1st}
\end{figure}
\begin{figure}[t]
    \centering  
    \includegraphics[clip,scale=\figscale, trim = 8 0 0 0]{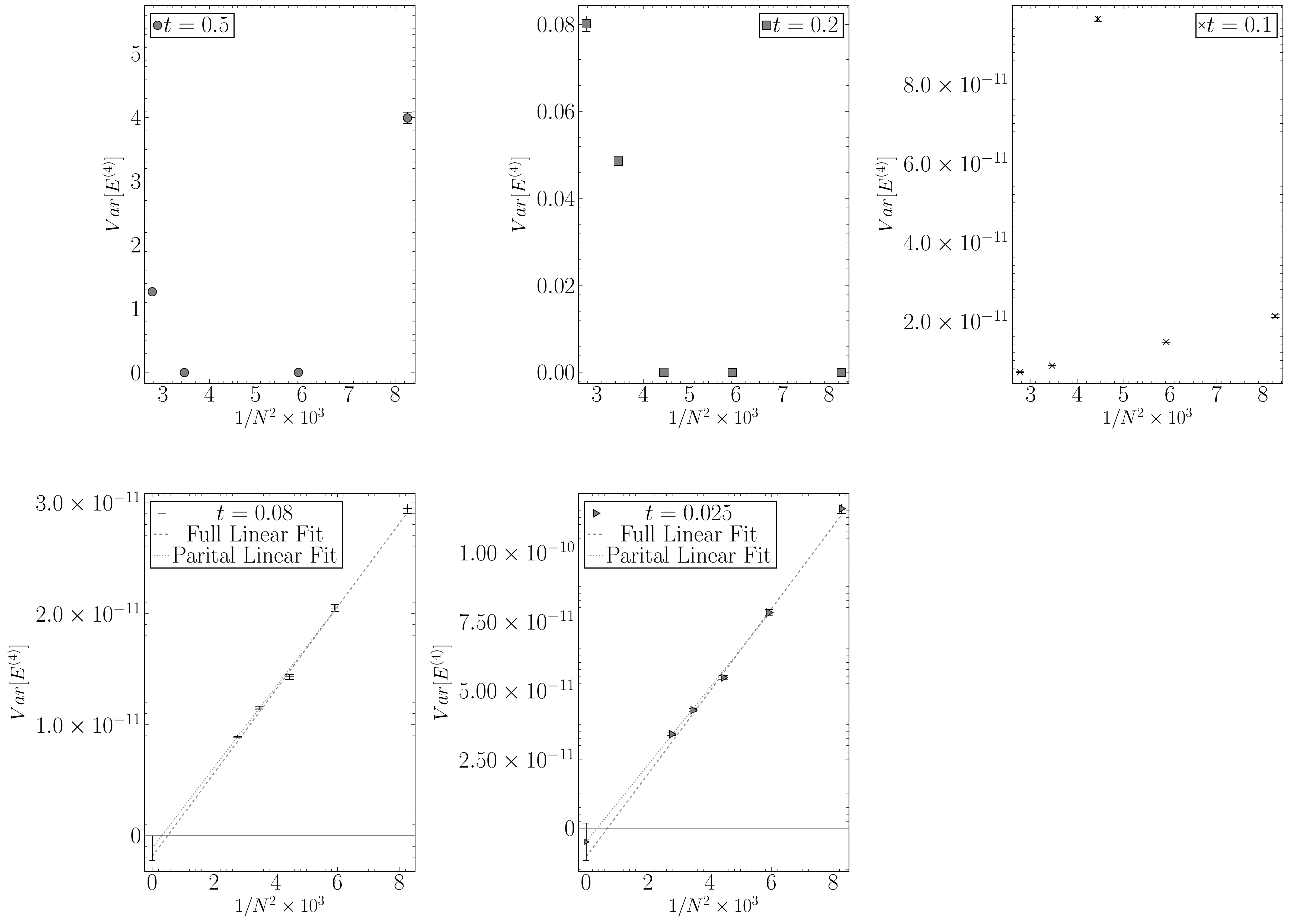}
    \caption{Same as Fig.~\ref{fig:hyper1st}, but for the fourth-order case.}
    \label{fig:hyper4th}
\end{figure}

Fig.~\ref{fig:hyper1st} shows the $\NC$ dependence of the variance of the leading order coefficient for each $t$.
The dashed and dotted lines show the linear fit in $1/\NC^2$ on the data with and without $N=11$, respectively.
The errors as $\NC\to\infty$ include the statistical error and systematic error assigned by fittings with and without the data at $\NC=11$.
Although the variances are small in $\order{10^{-6}}$--$\order{10^{-7}}$, those for $t>0.08$ show a non-smooth behavior.
While those for $t \le 0.08$ show a smooth  linear dependence resulting in the factorization property in the large $\NC$ limit.
Fig.~\ref{fig:hyper4th} shows those for the fourth order coefficient for each $t$. 
Similarly to Fig.~\ref{fig:hyper1st}, the variances for $t \le 0.08$ show a smooth  linear  dependence in the large $\NC$ limit.
The large $\NC$ factorization property is met, so our choice of $t=0.05$ is valid even though the reason for the irregular behavior in longer mean trajectory length $t > 0.08$ 
is not fully understood in this study.

\section{Fourth order coefficients of PCM}
\label{appendix:NSPTPCM}
The results of the lattice PCM's fourth-order coefficients evaluated with NSPT are included in this appendix.
Bruckmann and Puhr have been evaluated the coefficient of PCM up to $\order{\lambda^{20}}$\cite{Bruckmann:2019mky}
and their raw data are available in Ref.~\citen{BruckmannBuhrNSPTdata}. We perform an additional
NSPT simulation for the PCM with higher statistics than that of Refs.~\citen{Bruckmann:2019mky,BruckmannBuhrNSPTdata} in order to compare the PCM and the TRPCM at a comparable statistical error level for the first four order coefficients. In the range of $L=12-48$ for the lattice size, we use nine different sizes with periodic boundary conditions within the PCM.
We use the 4th order OMF integrator with the HMD-based NSPT lacking a randomly generated trajectory length.
We fix the trajectory length at $t=0.05$ and use several time steps to take the vanishing $\delta t$ limit.
We added a global chiral symmetry breaking term to the MD equation, similar to the gauge fixing term
and reunitarization process to stabilize the MD trajectory. 
We accumulated $\num{100000}$ trajectories for each simulation parameter, and computed the perturbation coefficients at $\NC=11$ and $19$.
The statistical errors are estimated using the single eliminated jackknife method after binning every 1000 trajectories.
After taking $\delta t\to 0$ limit with $\order{\delta t^4}$ scaling, 
the infinite volume limit is taken with the RG-based method
as described in Ref.~\citen{Bruckmann:2019mky}.
We include the perturbative beta function in the lattice scheme up to third order~\cite{weakcouplingPCM}  
and a linear term of $1/L^4$ in the RG-based fit function.

Table~\ref{tab:NSPT_FV_PCM} contains the coefficients for the first forth orders with the PCM at various lattice sizes $L$.
The $L\to \infty$ limit of $\expval{E^{(4)}}$ is extracted with the simultaneous fit using the RG-based volume dependence model 
with the constraint on the analytic values for $\expval{E^{(k=1,2,3)}}$ of the PCM in the $L\to \infty$ limit.
Figs.~\ref{fig:PCMinfVlimitN11} and \ref{fig:PCMinfVlimitN19} show the $1/L^2$ dependence and the fit results to the infinite volume limit
at $\NC=11$ and $19$, respectively.
The fourth-order coefficients are fitted while the first three order
coefficients are fixed to the analytical values. 
We observed a reasonable $\chi^2/\text{DoF}$ for the fitting.
Table~\ref{tab:PCM4loopNSPT} shows the fourth-order coefficients of the PCM in the infinite volume limit at $\NC=11$ and $19$. 

\renewcommand{\figscale}{0.45}
\begin{figure}[t]
    \centering  
    \includegraphics[clip,scale=\figscale,trim = 0 4 0 5]{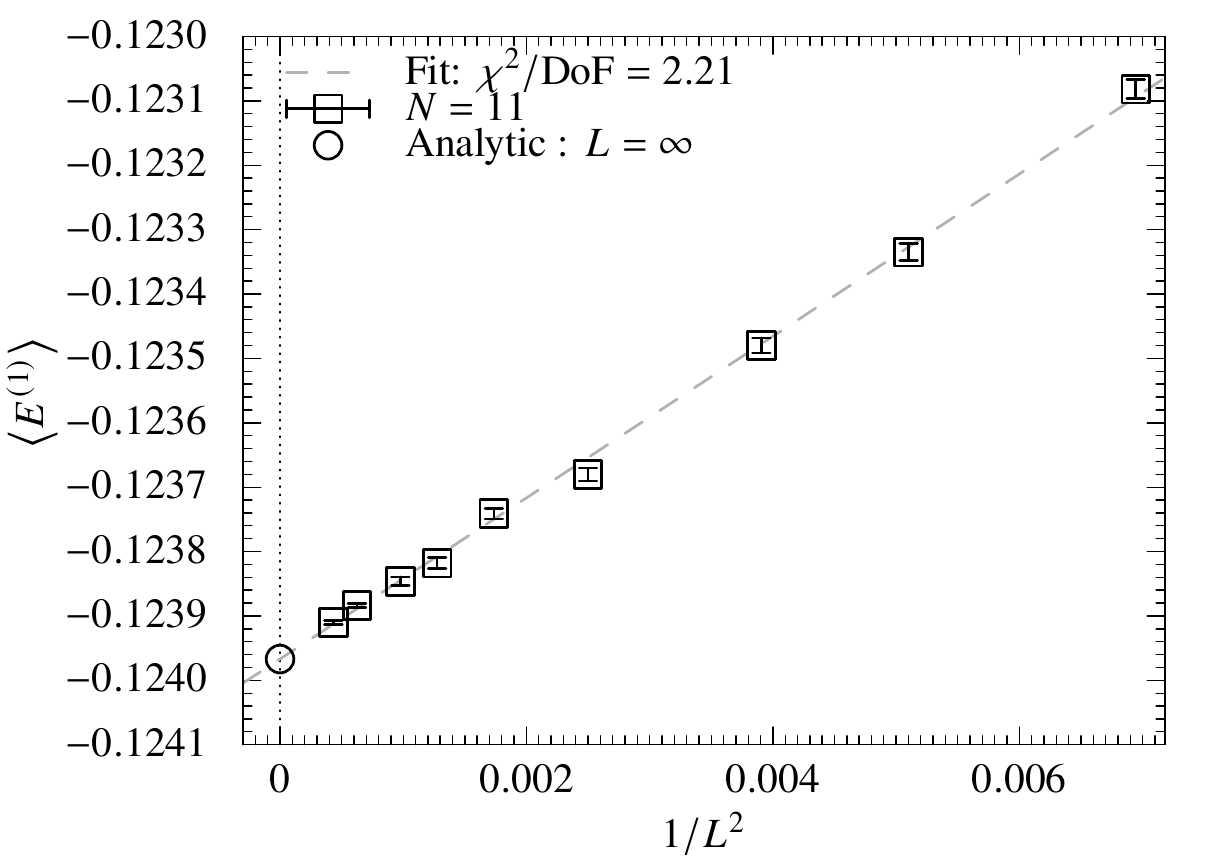}
    \includegraphics[clip,scale=\figscale,trim = 0 4 0 5]{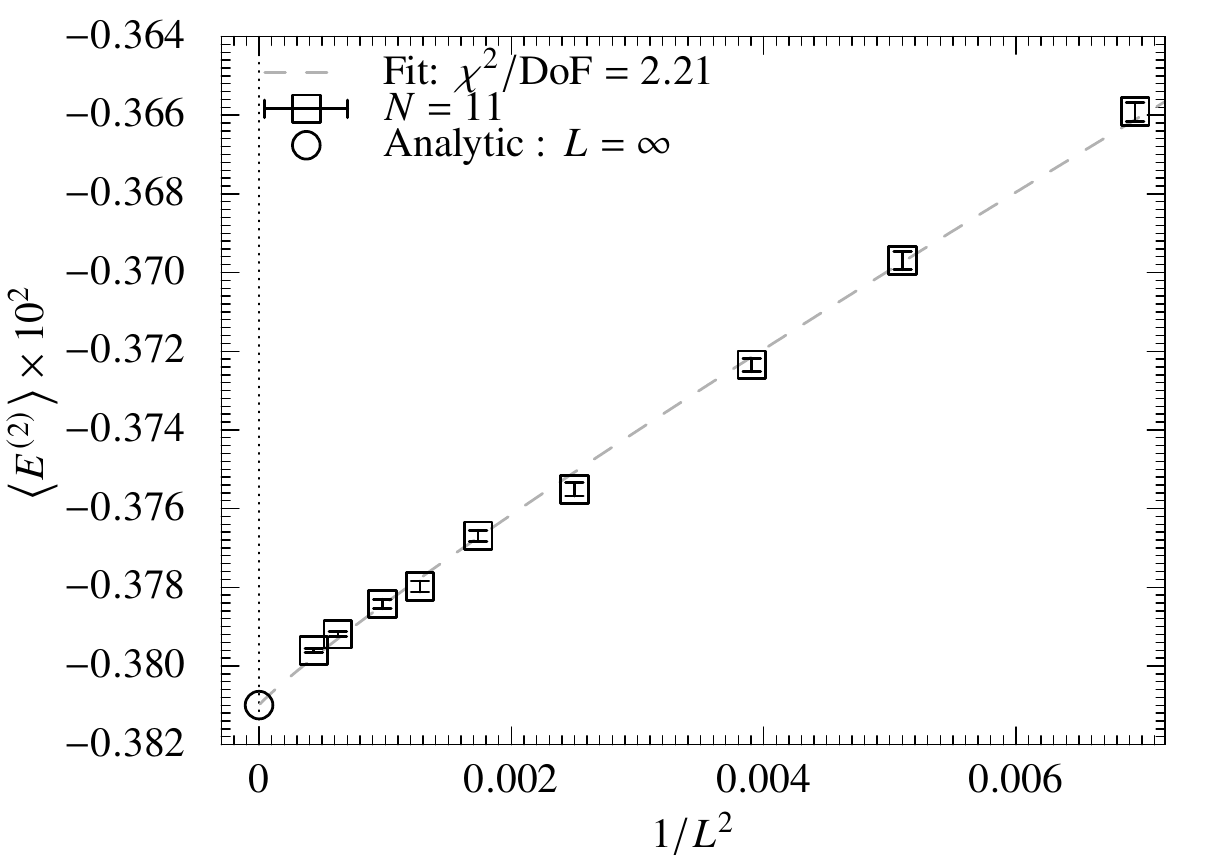}
    \includegraphics[clip,scale=\figscale,trim = 0 4 0 5]{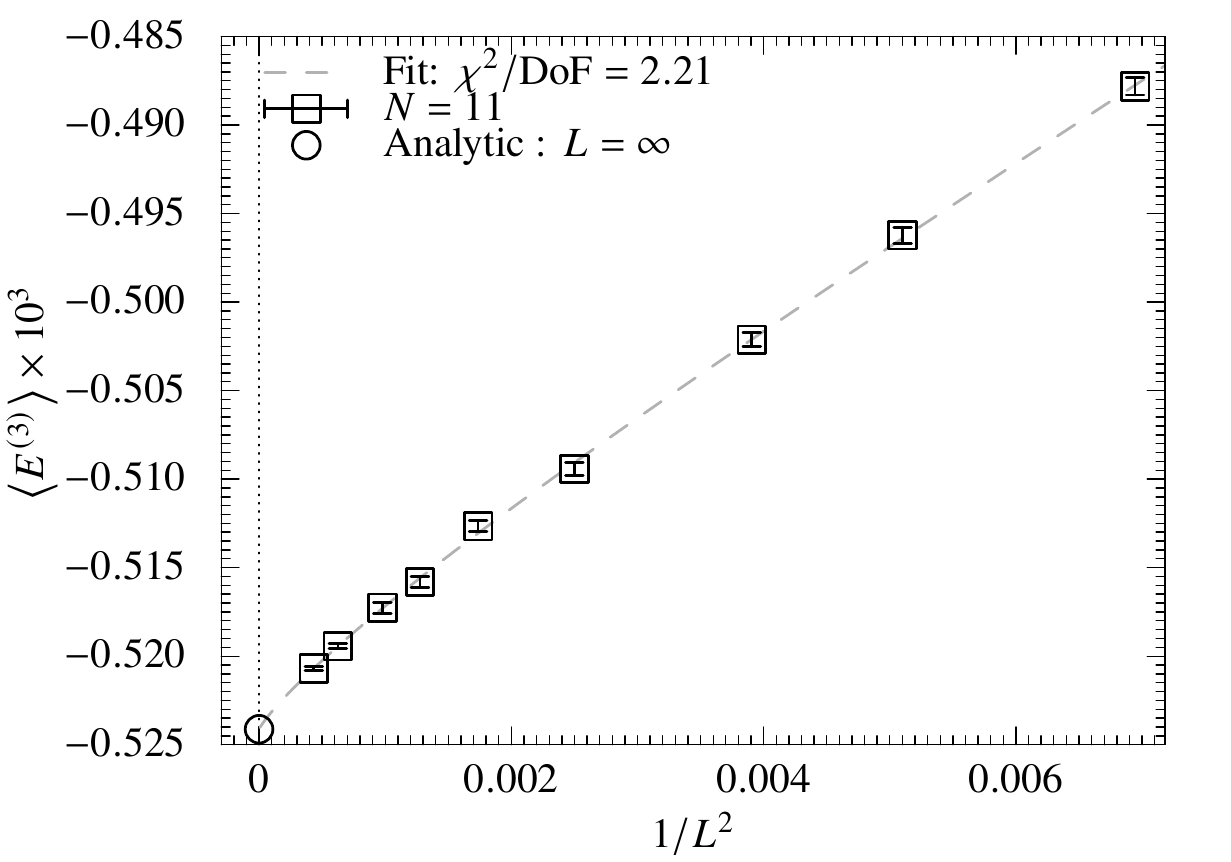}
    \includegraphics[clip,scale=\figscale,trim = 0 4 0 5]{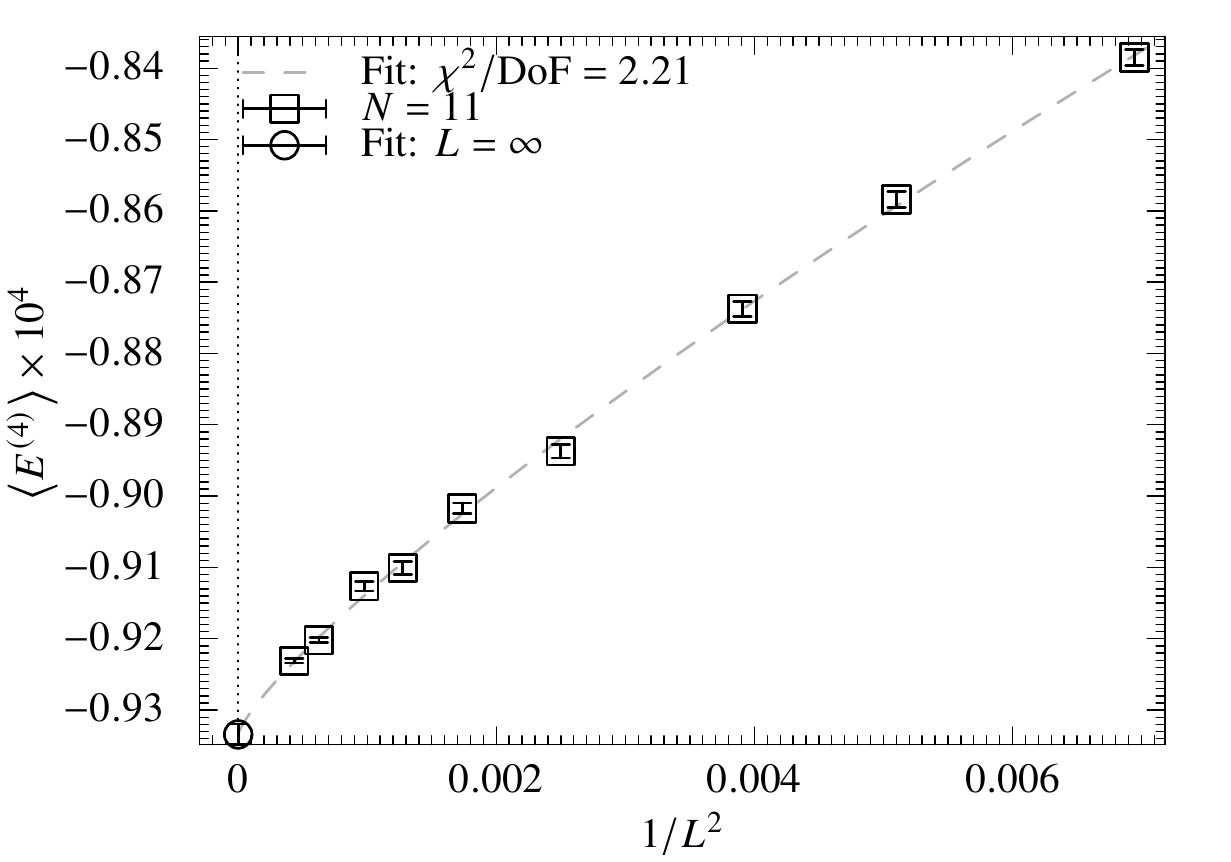}
    \caption{Volume dependence of $\expval{E^{(k)}}$ of the PCM at $\NC=11$.}
    \label{fig:PCMinfVlimitN11}
\end{figure}
\begin{figure}[t]
    \centering  
    \includegraphics[clip,scale=\figscale,trim = 0 4 0 5]{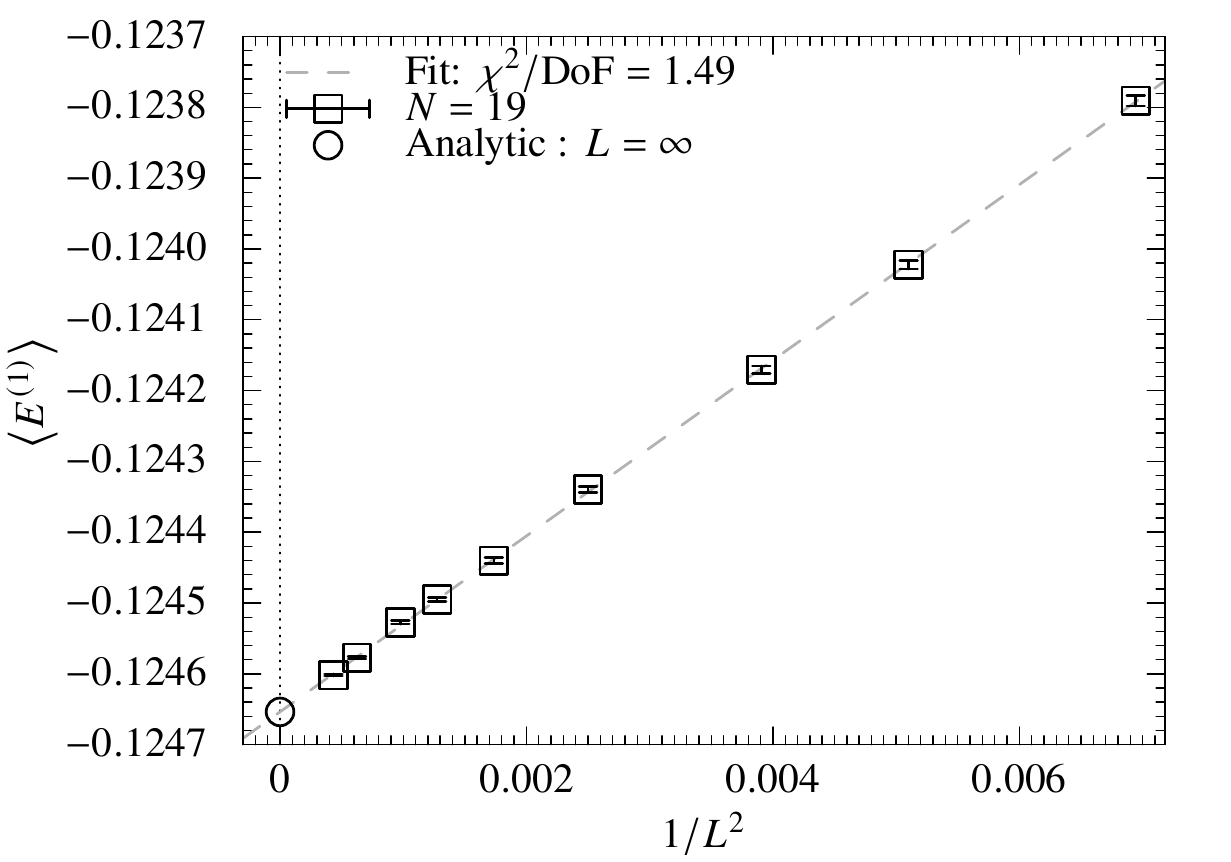}
    \includegraphics[clip,scale=\figscale,trim = 0 4 0 5]{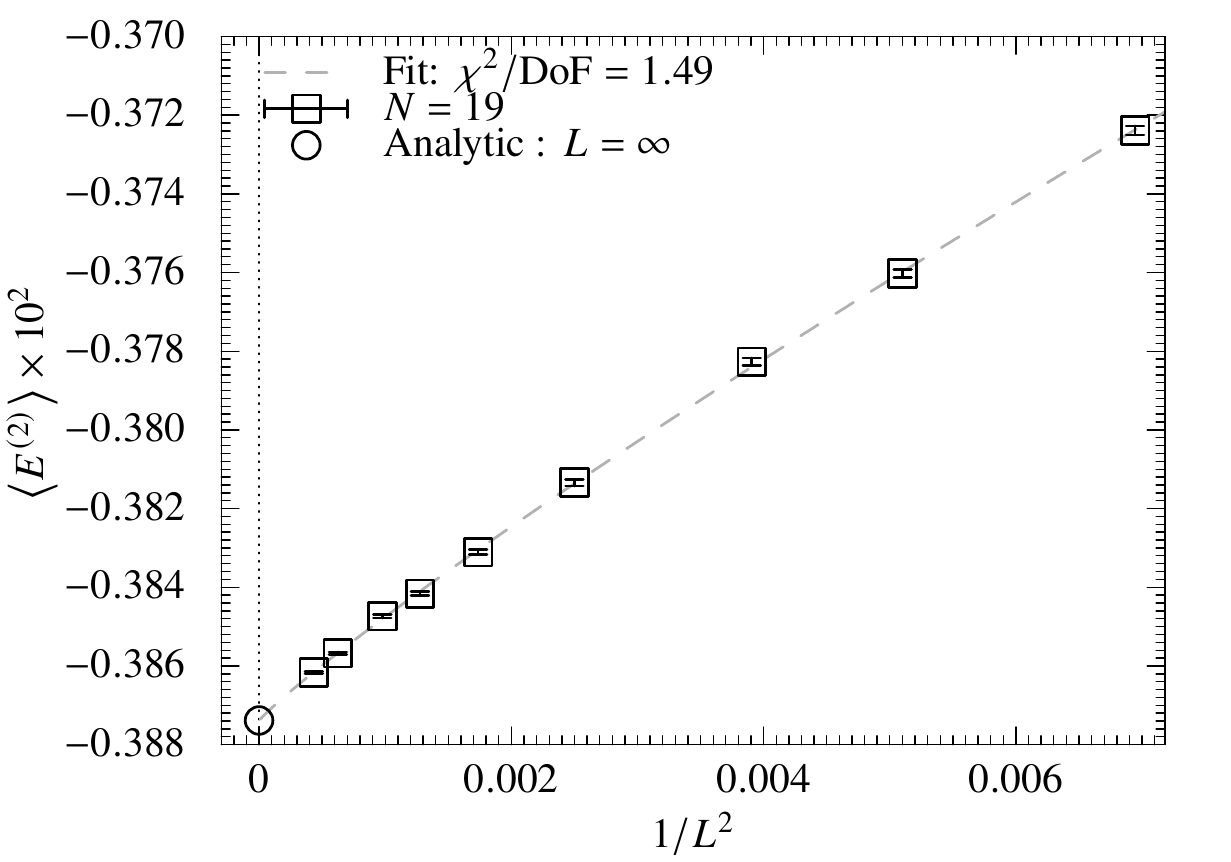}
    \includegraphics[clip,scale=\figscale,trim = 0 4 0 5]{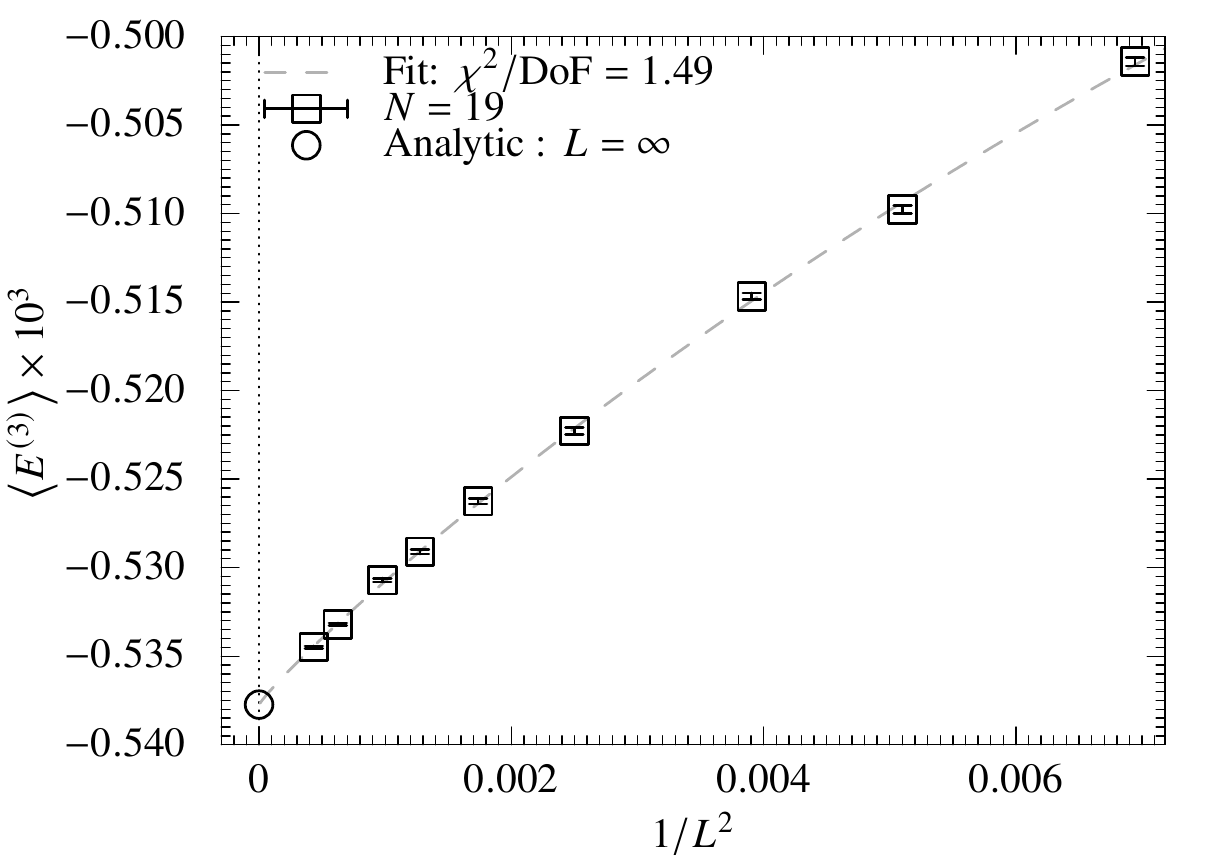}
    \includegraphics[clip,scale=\figscale,trim = 0 4 0 5]{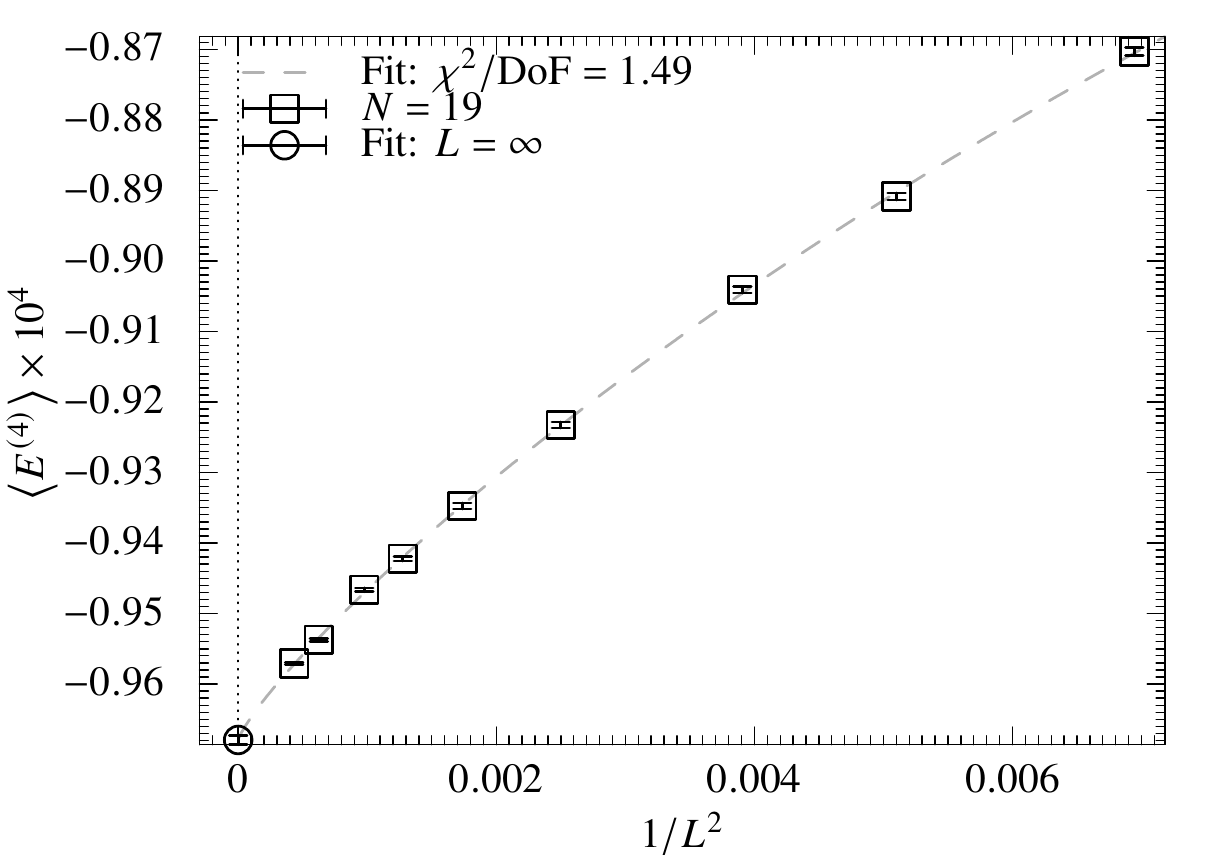}
    \caption{Volume dependence of $\expval{E^{(k)}}$ of the PCM at $\NC=19$.}
    \label{fig:PCMinfVlimitN19}
\end{figure}

%%%%%%%%%%%%%%%%%%%% Lattice PCM data (N=11,19) uses siunitx.sty ver 3 %%%%%%%%%%%%%%%
\begin{table}[t]
\tbl{Perturbative coefficients of the internal energy for the PCM ($\NC=11, 19$).}
{\begin{tabular}{cc%
                 S[round-mode = uncertainty, table-format=2.10(2)]%
                 S[round-mode = uncertainty, table-format=2.10(2)]%
                 S[round-mode = uncertainty, table-format=2.10(2)]%
                 S[round-mode = uncertainty, table-format=2.10(2)]}  \toprule
 {$\NC$} &
 {$L$} &
 {$\expval{E^{(1)}}$} &
 {$\expval{E^{(2)}}$} &
 {$\expval{E^{(3)}}$} &
 {$\expval{E^{(4)}}$} \\ \midrule
 11 & 12 & -0.123081576884( 0.000014838171) & -0.003659117385( 0.000002388210) & -0.000487811194( 0.000000489249) & -0.000083849059( 0.000000111748) \\
    & 14 & -0.123335047932( 0.000013014924) & -0.003696997447( 0.000002259652) & -0.000496222237( 0.000000454257) & -0.000085841338( 0.000000115521) \\
    & 16 & -0.123479859716( 0.000011782478) & -0.003723486854( 0.000001705552) & -0.000502126170( 0.000000397004) & -0.000087375236( 0.000000107112) \\
    & 20 & -0.123680507199( 0.000010053493) & -0.003755112264( 0.000001691481) & -0.000509426917( 0.000000367239) & -0.000089369911( 0.000000095097) \\
    & 24 & -0.123741104518( 0.000008146172) & -0.003766930524( 0.000001424986) & -0.000512658275( 0.000000310778) & -0.000090172261( 0.000000075161) \\
    & 28 & -0.123817912003( 0.000008177885) & -0.003779818359( 0.000001384803) & -0.000515813061( 0.000000306780) & -0.000091007510( 0.000000086885) \\
    & 32 & -0.123846254429( 0.000006612007) & -0.003784250968( 0.000001206373) & -0.000517270122( 0.000000305216) & -0.000091261047( 0.000000068771) \\
    & 40 & -0.123883456174( 0.000003243927) & -0.003791893924( 0.000000651007) & -0.000519431213( 0.000000142382) & -0.000092017695( 0.000000036337) \\
    & 48 & -0.123909982139( 0.000002840186) & -0.003796060343( 0.000000520305) & -0.000520694113( 0.000000115745) & -0.000092311566( 0.000000029710) \\
  \midrule
 19 & 12 & -0.123790798947( 0.000007321352) & -0.003723867294( 0.000001138531) & -0.000501417996( 0.000000245941) & -0.000087031548( 0.000000057756) \\
    & 14 & -0.124022434091( 0.000005908049) & -0.003760250434( 0.000000994926) & -0.000509770964( 0.000000210031) & -0.000089086623( 0.000000048705) \\
    & 16 & -0.124170557793( 0.000005187333) & -0.003782697663( 0.000000979573) & -0.000514681509( 0.000000191985) & -0.000090409206( 0.000000045063) \\
    & 20 & -0.124339728609( 0.000004391476) & -0.003813382941( 0.000000838315) & -0.000522282989( 0.000000180396) & -0.000092326289( 0.000000043149) \\
    & 24 & -0.124440221734( 0.000004083894) & -0.003831043947( 0.000000701909) & -0.000526241164( 0.000000159733) & -0.000093474680( 0.000000042828) \\
    & 28 & -0.124494650659( 0.000002695944) & -0.003841642681( 0.000000488305) & -0.000529110928( 0.000000119808) & -0.000094225526( 0.000000029801) \\
    & 32 & -0.124527305952( 0.000002436007) & -0.003847327097( 0.000000439050) & -0.000530702058( 0.000000106665) & -0.000094663749( 0.000000026738) \\
    & 40 & -0.124577313176( 0.000001613167) & -0.003856737147( 0.000000303401) & -0.000533202350( 0.000000071121) & -0.000095371609( 0.000000019576) \\
    & 48 & -0.124601698341( 0.000001352607) & -0.003861675896( 0.000000261684) & -0.000534485283( 0.000000063994) & -0.000095708476( 0.000000016443) \\
  \bottomrule
\end{tabular}
\label{tab:NSPT_FV_PCM}}
\end{table}
%%%%%%%%%%%%%%%%%%%%%%%%%%%%%%%%%%%%%%%%%%%%%%%%%%
\begin{table}[t]
\tbl{Fourth-order perturbative coefficients of the PCM in the infinite volume limit.}
{\begin{tabular}{c%
                S[round-mode = uncertainty,                 table-format=2.9(2)]%
                S[round-mode = places, round-precision = 2, table-format=2.2]}\toprule
 {$N$} &  {$\expval{E^{(4)}}$} & {$\chi^2/\text{DoF}$} \\\midrule
 11 & -0.000093340473( 0.000000144169) & 2.2085 \\
 19 & -0.000096792359( 0.000000067677) & 1.4869 \\
\bottomrule
\end{tabular}\label{tab:PCM4loopNSPT}}
\end{table}
%%%%%%%%%%%%%%%%%%%%%%%%%%%%%%%%%%%%%%%%%%%%%%%%%%%%%%%%%%%%%%%

\section{Next-to-leading order coefficient of the TRPCM using analytic method and comparison to NSPT}
\label{appendix:ananl}
Analytically, the standard perturbation method is used to calculate the internal energy's
next-to-leading order coefficient for the TRPCM. In this appendix, we compare the results between NSPT and the analytic formula in this appendix. 
The analytic formula for $\expval{E^{(2)}}_{\mathrm{ANA}}$ is
\begin{align}
  \expval{E^{(2)}}_{\mathrm{ANA}} &= E^{(2)}_{\mathrm{planar}} + E^{(2)}_{\mathrm{non-planar}},
\end{align}
\begin{align}
E^{(2)}_{\mathrm{planar}} &= \dfrac{1}{4} \qty[ \dfrac{1}{24\NC^4} \sum_{q}' \bm{P}(q)-\dfrac{1}{64}\qty(1-\dfrac{1}{\NC^2})^2],\\
E^{(2)}_{\mathrm{non-planar}} &= \dfrac{1}{16 \NC^4} \sum_{p}'\sum_{q}' \bm{P}(p)\bm{P}(q)Q(p,q)\cos\qty(\dfrac{\NC\bar{K}}{2\pi}\qty(p_1 q_2-p_2q_1)),
\end{align}
\begin{align}
\bm{P}(p) &= \dfrac{1}{2\sum_\mu\qty(1-\cos(p_\mu))},\\
Q(p,q) &= \sum_{\mu}\qty(-\dfrac{1}{6}+\dfrac{2}{3}\cos(p_{\mu})-\dfrac{1}{2}\cos(p_{\mu}+q_{\mu})).
\end{align}
The primed sum $\sum'_{p}$ means that $p=0$ is excluded from the sum and $p_{\mu} = 2\pi n_{\mu}/\NC, n_{\mu} = 0,\dots, \NC-1$.

Fig.~\ref{fig:diff} shows the difference $\expval{E^{(2)}}_{\mathrm{NSPT}}-\expval{E^{(2)}}_{\mathrm{ANA}}$,
where $\expval{E^{(2)}}_{\mathrm{NSPT}}$ and $\expval{E^{(2)}}_{\mathrm{ANA}}$ are the results with NSPT and standard perturbation theory, respectively.
They are consistent within the statistical error.

\renewcommand{\figscale}{0.50}
\begin{figure}[t]
\centering
\includegraphics[clip,scale=\figscale,trim=0 4 0 4]{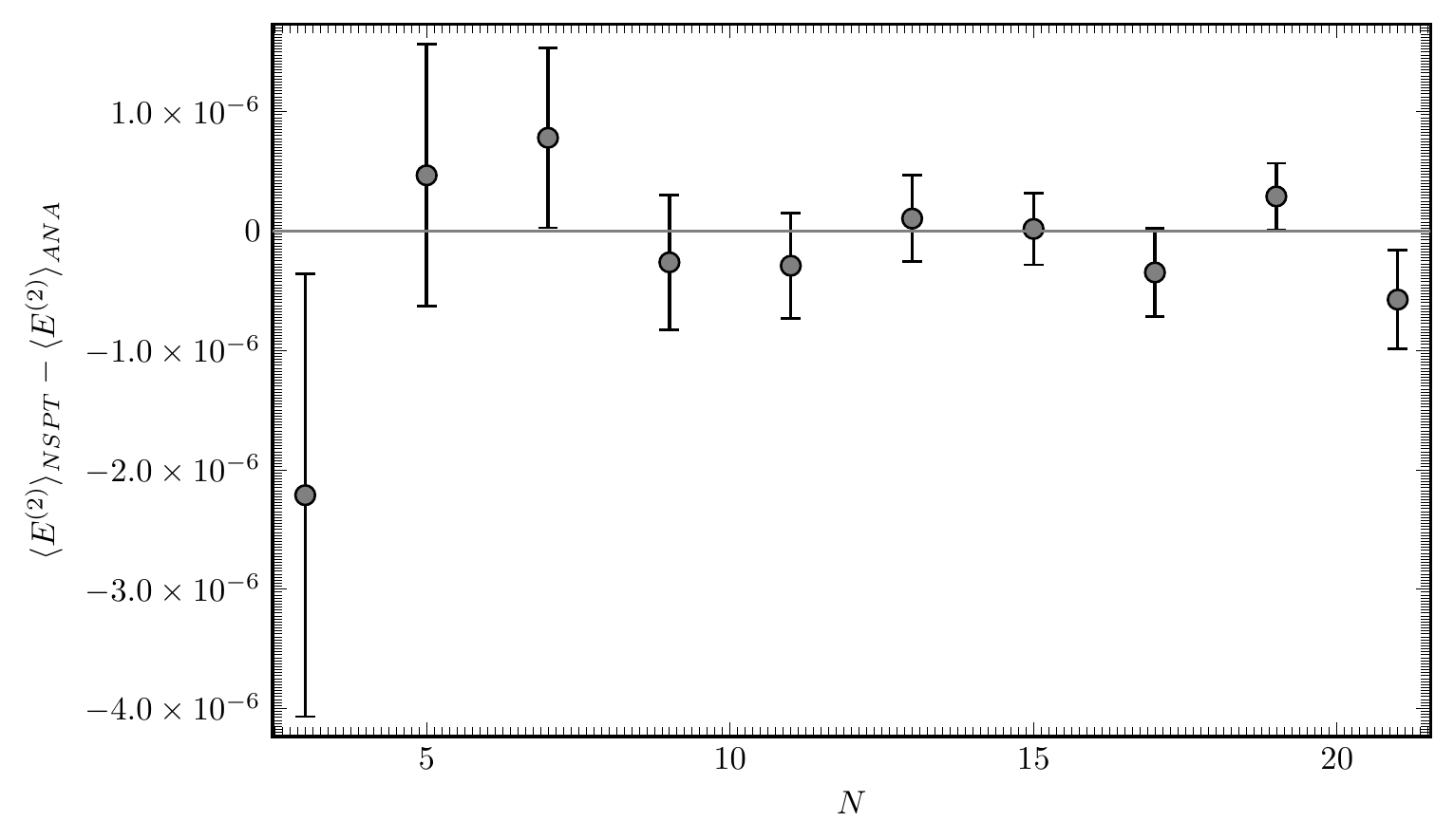}
\caption{Difference between $\expval{E^{(2)}}_{\mathrm{NSPT}}$ and $\expval{E^{(2)}}_{\mathrm{ANA}}$ in next-to-leading order level.}
 \label{fig:diff}
\end{figure}

%\end{appendices}
%\newpage
%\end{appendices}
  
%\begin{thebibliography}{000} %for 3 digits
%\begin{thebibliography}{00}  %for 2 digits
%\begin{thebibliography}{0}   %for 1 digit
\bibliographystyle{ws-ijmpa}
\bibliography{nspt_trpcm_refs.bib}
%\end{thebibliography}
\end{document}